\documentclass[a4paper,11pt]{article}
\pdfoutput=1 
\usepackage{jcappub} 
\usepackage{graphicx}
\usepackage{xcolor}
\usepackage[caption=false]{subfig}
\usepackage{mathrsfs,mathtools}
\usepackage{physics,amssymb}
\usepackage{siunitx}
\usepackage{bm}
\usepackage{braket}
\usepackage{listings}
\usepackage{cases}
\usepackage{comment}
\usepackage{soul}
\usepackage{cancel}
\usepackage{cases}
\usepackage[utf8]{inputenc}
\usepackage{url}
\usepackage{longtable}
\usepackage{xspace}
\usepackage{acronym}
\usepackage{aas_macros}

\newcommand{\ee}{\mathrm{e}}

\DeclareMathOperator{\Erf}{Erf}
\newcommand{\Mpl}{M_\mathrm{Pl}}
\newcommand{\ns}{n_{\mathrm{s}}}

\newcommand{\FPT}{\mathrm{FPT}}
\newcommand{\FP}{\mathrm{FP}}
\newcommand{\bw}{\mathrm{bw}}

\newcommand{\CC}{C\nolinebreak[4]\hspace{-.05em}\raisebox{.4ex}{\tiny\bf ++}\xspace}

\newcommand{\uc}{\mathrm{c}}

\newcommand{\uf}{\mathrm{f}}

\newcommand{\calL}{\mathcal{L}}

\newcommand{\calN}{\mathcal{N}}
\newcommand{\calO}{\mathcal{O}}

\newcommand{\calP}{\mathcal{P}}

\newcommand{\calNhat}{\mathcal{N}}
\newcommand{\Ne}{N}
\newcommand{\D}{\mathscr{N}}
\newcommand{\HI}{H_{\rm inf}}

\newcommand{\beae}[1]{\begin{equation}\begin{aligned} #1 \end{aligned}\end{equation}}

\newcommand{\bae}[1]{\begin{align} #1 \end{align}}
\newcommand{\bce}[1]{\begin{cases} #1 \end{cases}}
\newcommand{\dps}{\displaystyle}

\newcommand{\bme}[1]{\begin{multline} #1 \end{multline}}

\newcommand{\bmbe}[1]{\begin{multlined}[b] #1 \end{multlined}}

\definecolor{MONZA}{HTML}{CF000F}
\definecolor{DARKBLUE}{HTML}{00008b}
\definecolor{DARKMAGENTA}{HTML}{8b008b}

\def\beqa{\begin{eqnarray}}
\def\eeqa{\end{eqnarray}}

\newcommand{\GeV}{\,\si{GeV}}
\newcommand{\MeV}{\,\si{MeV}}

\def\lmk{\left(}
\def\rmk{\right)}
\def\lkk{\left[}
\def\rkk{\right]}
\def\dd{{\rm d}}

\newcommand{\eq}[1]{Eq.~(\ref{#1})}

\newcommand{\beq}{\begin{eqnarray}} 
\newcommand{\eeq}{\end{eqnarray}}

\newcommand{\infl}{\zeta}

\newcommand{\curv}{\mathcal{R}}
\newcommand{\Nc}{\Ne_{\rm c}}
\newcommand{\Nend}{\Ne_\mathrm{end}}

\begin{document}
\rightline{TU-1196}

\title{Stochastic dynamics of multi-waterfall hybrid inflation and formation of primordial black holes}
\date{\today}

\author[a,b]{Yuichiro Tada}
\author[c,d]{and Masaki Yamada}
\affiliation[a]{
Institute for Advanced Research, Nagoya University, \\
Furo-cho Chikusa-ku, 
Nagoya 464-8601, Japan
} 
\affiliation[b]{
Department of Physics, Nagoya University, 
Furo-cho Chikusa-ku,
Nagoya 464-8602, Japan
} 
\affiliation[c]{
Department of Physics, Tohoku University, 
Sendai, Miyagi 980-8578, Japan}  
\affiliation[b]{
Frontier Research Institute for Interdisciplinary Sciences, Tohoku University, \\
Sendai, Miyagi 980-8578, Japan}  

\emailAdd{tada.yuichiro.y8@f.mail.nagoya-u.ac.jp}
\emailAdd{m.yamada@tohoku.ac.jp}

\abstract{
We show that a hybrid inflation model with multiple waterfall fields can result in the formation of \acp{PBH} with an astrophysical size, by using an advanced algorithm to follow the stochastic dynamics of the waterfall fields. This is in contrast to the case with a single waterfall field, where the wavelength of density perturbations is usually too short to form PBHs of the astrophysical scale (or otherwise \acp{PBH} are overproduced and the model is ruled out) unless the inflaton potential is tuned. In particular, we demonstrate that PBHs with masses of order $10^{20}\, {\rm g}$ can form after hybrid inflation consistently with other cosmological observations if the number of waterfall fields is about $5$ for the case of instantaneous reheating. Observable gravitational waves are produced from the second-order effect of large curvature perturbations as well as from the dynamics of texture or global defects that form after the waterfall phase transition. 
}

\maketitle

\acrodef{CMB}{cosmic microwave background}
\acrodef{BH}{black hole}
\acrodef{PBH}{primordial black hole}
\acrodef{SMBH}{supermassive black hole}
\acrodef{DM}{dark matter}
\acrodef{GW}{gravitational wave}
\acrodef{FP}{Fokker--Planck}
\acrodef{SSB}{spontaneous symmetry breaking}
\acrodef{NG}{Nambu--Goldstone}

\acresetall

\section{Introduction}

The seeds of cosmological large-scale structures are generated by curvature perturbations during inflation, and their signals are imprinted and observed as the temperature anisotropies of the \ac{CMB}. 
The spectrum of curvature perturbations is not exactly flat, which is consistent with the prediction of slow-roll inflationary scenario~\cite{Planck:2018vyg}. 
However, there are several puzzles in the formation of small-scale structures, such as the formation of \acp{SMBH}~\cite{LyndenBell:1969yx, Kormendy:1995er} and \acp{BH} observed by the LIGO--Virgo--KAGRA colllaboration~\cite{LIGOScientific:2016dsl, LIGOScientific:2021djp}. 
Although their evolution is highly non-linear, it is said that their number densities are in tension with the standard scenario of BH formation in astrophysics. 
One therefore needs to consider a primordial origin of such BHs 
called \acp{PBH}~\cite{Hawking:1971ei,Carr:1974nx,Carr:1975qj} (see also the recent review article on \ac{PBH}~\cite{Escriva:2022duf}).
It is also discussed that astrophysical size PBHs of order $10^{17\text{--}23} \, {\rm g}$ can explain the \ac{DM} in the Universe~\cite{Chapline:1975ojl} (see, e.g., Refs.~\cite{Carr:2016drx,Inomata:2017okj,Inomata:2017vxo,Carr:2020gox}). 
Among several scenarios of the formation of PBHs, 
it would be minimal to consider the formation of those small-scale structures by generating large curvature perturbations in the small scales during inflation (see, e.g., Refs.~\cite{Ivanov:1994pa, GarciaBellido:1996qt, Kawasaki:1997ju, Yokoyama:1998pt, Garcia-Bellido:2017mdw, Hertzberg:2017dkh} and also references in Ref.~\cite{Escriva:2022duf}).%
\footnote{
For other scenarios for PBH formation, see scenarios with cosmic strings~\cite{Hawking:1987bn, Garriga:1992nm, Caldwell:1995fu}, bubble collisions~\cite{Hawking:1982ga}, domain walls~\cite{Garriga:1992nm, Khlopov:2008qy, Garriga:2015fdk, Deng:2016vzb}, and collapse of vacuum bubbles~\cite{Garriga:2015fdk, Deng:2017uwc, Deng:2018cxb,Deng:2020mds}, etc. 
}
The curvature perturbations are converted to density perturbations after inflation, and the dense regions collapse to form PBHs when the corresponding wavelength enters the horizon~\cite{Hawking:1971ei,Carr:1974nx,Carr:1975qj}.

To generate large curvature perturbations, a bump or at least a very flat region for inflaton potential is required~\cite{GarciaBellido:1996qt,Kawasaki:1997ju,Yokoyama:1998pt,Kawasaki:2006zv,Kawaguchi:2007fz,Kohri:2007qn,Frampton:2010sw,Drees:2011yz,
Kawasaki:2012kn,Kawasaki:2012wr,Kawasaki:2016pql,Inomata:2016rbd,Ezquiaga:2017fvi,Kannike:2017bxn,Germani:2017bcs,Motohashi:2017kbs,Ballesteros:2017fsr,Hertzberg:2017dkh,Cicoli:2018asa,Cheong:2019vzl,Ballesteros:2020qam,Pi:2021dft,Geller:2022nkr}. 
This can be naturally realised by hybrid inflation models because a waterfall field has a vanishing mass at the waterfall phase transition~\cite{Linde:1993cn}. 
Remarkably, it can generate large density perturbations at small scales that result in the formation of PBHs~\cite{Clesse:2010iz,Kodama:2011vs,Mulryne:2011ni}. 
Since large density perturbations are generated during the waterfall phase transition, the waterfall field experiences the stochastic dynamics that requires detailed numerical simulations (see Refs.~\cite{Starobinsky:1982ee,Starobinsky:1986fx,Nambu:1987ef,Nambu:1988je,Kandrup:1988sc,Nakao:1988yi,Nambu:1989uf,Mollerach:1990zf,Linde:1993xx,Starobinsky:1994bd} for the first papers on the subject) and was omitted in some literature. 
Several studies on hybrid inflation models 
reveal that 
the stochastic effect on the waterfall fields significantly alters the predictions of hybrid inflation models (see, e.g., Ref.~\cite{Martin:2011ib}). 
Ref.~\cite{Clesse:2015wea} provided an analytic method to calculate the spectrum of curvature perturbations in a model with a single waterfall field, considering  the stochastic effects under some approximations. Their analytic formula is partially consistent with the numerical simulations for the stochastic dynamics performed in Ref.~\cite{Kawasaki:2015ppx}. These studies are performed in models with a single waterfall fields, and we thereby concluded that the peak amplitude of curvature perturbations is too large to provide consistent astrophysical consequences even though the stochastic effect is considered.%
\footnote{
The resulting spectrum of curvature perturbations calculated in Ref.~\cite{Clesse:2015wea} indicates that the peak amplitude and corresponding peak frequency have a one-to-one correspondence without any free parameters. One can conclude that the mass scale of PBHs, if a significant amount can form, is much smaller than the scale of interest for astrophysics. They however applied their analytic results to PBH formations with errors. 
}
Our recent work~\cite{Tada:2023pue} shows that this problem can be evaded by considering a higher-order term for the inflaton potential, though it requires a tuning in a parameter. 
As another aspect, the hybrid inflation model with a single waterfall field suffers from the so-called domain-wall problem because the $Z_2$ symmetry is spontaneously broken after the waterfall phase transition. 
This problem can be evaded by introducing a small explicit $Z_2$-symmetry breaking term, which however may affect the dynamics of waterfall phase transition (see, e.g., Ref.~\cite{Braglia:2022phb}).

In this paper, we analyze a hybrid inflation model with multiple waterfall fields, considering the stochastic effect on the waterfall fields. 
They are assumed to be symmetric under O($\D$) symmetry. 
The idea to introduce multiple waterfall fields was considered in Ref.~\cite{Halpern:2014mca}, although they did not consider the 
stochastic dynamics of waterfall fields. 
We show that the spectrum of curvature perturbations can be reduced with a fixed peak frequency for this case. 
Alternatively, the peak frequency can be reduced for a fixed peak amplitude of curvature perturbations. 
This can result in the formation of PBHs at astrophysically interesting mass scales when the number of waterfall fields is sufficiently large. 
We employ a sophisticated numerical method proposed and developed in Refs.~\cite{Fujita:2013cna,Fujita:2014tja,Vennin:2015hra,Ando:2020fjm,Tada:2021zzj} to incorporate the stochastic dynamics of multiple waterfall fields. We also confirm that the numerical results can be qualitatively understood by the analytical calculation of Ref.~\cite{Clesse:2015wea}, extending it for the case with multiple waterfall fields. In particular, the peak amplitude of curvature perturbations is proportional to the inverse of the number of waterfall fields. 
We provide a semi-analytic formula for the spectrum of curvature perturbations, which is useful to consider the formation of PBHs. From the formula, we particularly demonstrate that PBHs with mass of order $10^{20}\, {\rm g}$ can form after hybrid inflation consistently with other cosmological observations 
if the number of waterfall fields is about $4$ or $5$ for the case of instantaneous reheating. 
Moreover, 
the domain wall problem is absent for the case with O($\D$) symmetric waterfall fields. 
The spontaneous symmetry breaking of O($\D$) symmetry after the waterfall phase transition results in the formation of texture (for $\D = 4$) or global defects (for $\D \ge 5$). 
\Acp{GW} are generated via the dynamics of those structure as well as the second-order effect of large curvature perturbations. We demonstrate that the GW signals can be observed by future GW experiments, including LISA.

The rest of the paper is organised as follows.
In Sec.~\ref{sec:formalism}, we first summarise the formalism for the stochastic equations for O($\D$) symmetric fields and write down the Fokker--Planck and Langevin equations. 
These equations are solved analytically and numerically in Sec.~\ref{sec:hybrid}, in hybrid inflation models with multiple waterfall fields. 
We provide a semi-analytic formula for the spectrum of curvature perturbations, where numerical coefficients are fitted by the results of numerical calculations. 
Furthermore, in Sec.~\ref{sec:parameters} 
we use it to consider the formation of PBHs and the prediction of GW spectrum. 
Sec.~\ref{sec:conclusion} is devoted to discussion and conclusions.

\section{Stochastic equations for symmetric fields}
\label{sec:formalism}

Let us first see the \ac{FP} and Langevin equations that describe the stochastic dynamics of O($\D$) symmetric fields during inflation. One will find that the system reduces to the one only for the radial mode in the slow-roll limit with an effective centrifugal force proportional to the number $\D$ of the fields due to the diffusion term. 
If the $\D$ fields correspond to the waterfall fields in the hybrid inflation, such a centrifugal force prevent them from staying around 
the origin of the potential and accordingly suppresses the curvature perturbation as we will see.

We employ the slow-roll approximation throughout this paper. 
The Langevin equation for $\D$ (canonical) inflaton fields $\psi_i$ ($i=1,2,\cdots,\D$) (which we will identify as waterfall fields in the context of hybrid inflation model) is expressed as (see, e.g., Ref.~\cite{Vennin:2015hra})
\bae{
    \partial_\Ne\psi_i=-\Mpl^2\frac{V_i}{V}+\frac{1}{2\pi}\sqrt{\frac{V}{3\Mpl^2}}\xi_i(\Ne),
}
where $\Ne$ is the e-folding number as the time variable, $\Mpl$ is the reduced Planck mass, $V$ is the inflatons' potential, $V_i=\partial_iV=\partial_{\psi_i}V$ is its first derivative, 
and $\xi_i(\Ne)$ is an independent stochastic noise:
\bae{
    \braket{\xi_i(\Ne)\xi_j(\Ne^\prime)}=\delta_{ij}\delta(\Ne-\Ne^\prime).
}
The corresponding \ac{FP} equation for the inflatons' probability distribution $P(\bm{\psi}=\Bqty{\psi_1,\psi_2,\cdots} \mid \Ne)$ is expressed as
\bae{
    \partial_\Ne P(\bm{\psi}\mid \Ne )=\partial_i\pqty{\Mpl^2\frac{V_i}{V}P(\bm{\psi}\mid \Ne )}+\frac{1}{2}\partial_i^2\pqty{\frac{V}{12\pi^2\Mpl^2}P(\bm{\psi}\mid \Ne)}.
}
This is equivalent to the following adjoint \ac{FP} equation for the distribution $P_\FPT(\calNhat\mid\bm{\psi})$ of the first passage time $\calNhat$ from a certain field space point $\bm{\psi}$ to the 
end-of-inflation surface: 
\bae{
    \partial_\calNhat P_\FPT(\calNhat\mid\bm{\psi})=-\Mpl^2\frac{V_i}{V}\partial_iP_\FPT(\calNhat\mid\bm{\psi})+\frac{1}{2}\frac{V}{12\pi^2\Mpl^2}\partial_i^2P_\FPT(\calNhat\mid\bm{\psi}).
}
Note that the forward e-folding number $\Ne$ is a deterministic time variable while the backward one $\calNhat$ is stochastic as it depends on the realisation of the inflatons' stochastic dynamics.

Let us then suppose that the inflatons are symmetric and the potential (and hence all physical variables in the slow-roll limit) depends only on the radius $\psi_r=\sqrt{\psi_i^2}$. In this case, $P_\FPT$ should solely be a function of $\calNhat$ and $\psi_r$. Therefore, the adjoint \ac{FP} equation reduces to (refer also to Ref.~\cite{Assadullahi:2016gkk})
\bae{
    \partial_\calNhat P_\FPT(\calNhat\mid \psi_r)=\bqty{\pqty{-\Mpl^2\frac{V_{\psi_r}}{V}+\frac{1}{2}\frac{V}{12\pi^2\Mpl^2}\frac{\D-1}{\psi_r}}\partial_{\psi_r}+\frac{1}{2}\frac{V}{12\pi^2\Mpl^2}\partial_{\psi_r}^2}P_\FPT(\calNhat\mid \psi_r).
}
One finds a new term $\propto V\frac{\D-1}{\psi_r}\partial_{\psi_r}$ from the Laplacian $V\partial_i^2$.
It can be understood as an effective single field system with a new centrifugal force dictated by the \ac{FP} equation
\bme{
    \partial_\Ne P(\psi_r\mid \Ne)=\partial_{\psi_r}\bqty{\pqty{\Mpl^2\frac{V_{\psi_r}}{V}-\frac{1}{2}\frac{V}{12\pi^2\Mpl^2}\frac{\D-1}{\psi_r}}P(\psi_r\mid \Ne)} \\ 
    +\frac{1}{2}\partial_{\psi_r}^2\pqty{\frac{V}{12\pi^2\Mpl^2}P(\psi_r\mid \Ne)},
    \label{eq:FP1}
}
or the corresponding Langevin equation
\bae{
    \partial_\Ne \psi_r=-\Mpl^2\frac{V_{\psi_r}}{V}+\frac{1}{2}\frac{V}{12\pi^2\Mpl^2}\frac{\D-1}{\psi_r}+\frac{1}{2\pi}\sqrt{\frac{V}{3\Mpl^2}}\xi_{\psi_r}(\Ne),
    \label{eq:langevin1}
}
with a normalised noise
\bae{
    \braket{\xi_{\psi_r}(\Ne)\xi_{\psi_r}(\Ne^\prime)}=\delta(\Ne-\Ne^\prime).
}
We below see the effect of this centrifugal force in the hybrid inflation model.

\section{Multi-waterfall hybrid inflation}
\label{sec:hybrid} 

In order for a general discussion, we consider the $(1+\D)$-field hybrid inflation with a phenomenological potential~\cite{Clesse:2015wea,Kawasaki:2015ppx}
\bae{
    V(\phi,\bm{\psi})=\Lambda^4\bqty{\pqty{1-\frac{\psi_i^2}{M^2}}^2+2\frac{\phi^2\psi_i^2}{\phi_\uc^2M^2}+\frac{\phi-\phi_\uc}{\mu_1}-\frac{(\phi-\phi_\uc)^2}{\mu_2^2}}.
    \label{eq:potential}
}
characterised by the five dimensionful parameters $\Lambda$, $M$, $\phi_\uc$, $\mu_1$, and $\mu_2$. 
Here and hereafter, the summation in terms of $i$ is implicit for $\psi_i^2$ ($\equiv \sum_i \psi_i^2$).
The inflaton potential is expanded around the critical point $\phi_\uc$ and the higher-order terms than the quadratic one are neglected as we are mainly interested in the dynamics around the waterfall phase transition. 
Among the inflatons $\phi$ and $\psi_i$ ($i=1,2,\cdots,\D$), the latter is often called \emph{waterfall} fields 
as they terminate inflation by their tachyonic instability in the standard scenario. 
For a large $\phi$, the waterfall fields stay around the origin 
and inflation occurs by the potential energy ($\simeq \Lambda^4$). The Hubble parameter during inflation is given by $\HI \simeq \Lambda^2/(\sqrt{3}\Mpl)$. 
When the field $\phi$ reaches the critical point $\phi_\uc$, 
$\psi_i$ becomes tachyonic (called the waterfall phase) and starts to roll down toward the potential minimum $\psi_i^2=M^2$. 
Depending on the parameters, the waterfall phase does not necessarily end quickly but can last as a second phase of inflation. 
The inflation ends when the slow-roll condition is violated. 
It is often controlled by the second slow-roll parameter along the waterfall direction $\eta_{ij}\equiv\Mpl^2\frac{V_{\psi_i\psi_j}}{V}$.
In particular, the potential of interest is symmetric with respect to $\psi_i$ and therefore the model is effectively a two-field system of $\phi$ and $\psi_r=\sqrt{\psi_i^2}$ with the centrifugal force disucssed around \eq{eq:langevin1} in the slow-roll limit. 
We below study this two-field system.

\subsection{Analytic estimations}
\label{sec:analytic}

The spectrum of curvature perturbations can be estimated by 
an analytic method proposed in Ref.~\cite{Clesse:2015wea} (refer also to Refs.~\cite{Kawasaki:2015ppx,Tada:2023pue}) for the hybrid inflation model. 
We extend their analysis to the case with multiple waterfall fields, including the leading-order effect of quadratic term for the inflaton potential. 

We first estimate the distribution of the waterfall fields at the critical point, which determines the dynamics in the waterfall phase and then the final curvature perturbation.
It follows the Langevin equations, 
\bae{
    \bce{
        \dps
        \partial_\Ne \phi=-\Mpl^2\frac{V_\phi}{V}+\frac{1}{2\pi}\sqrt{\frac{V}{3\Mpl^2}}\xi_\phi(\Ne), \\
        \dps
        \partial_\Ne \psi_r=-\Mpl^2\frac{V_{\psi_r}}{V}+\frac{1}{2}\frac{V}{12\pi^2\Mpl^2}\frac{\D-1}{\psi_r}+\frac{1}{2\pi}\sqrt{\frac{V}{3\Mpl^2}}\xi_{\psi_r}(\Ne),
    }
    \label{eq:stochastic_r}
}
with the independent noise
\bae{
\label{eq:noise} \braket{\xi_\phi(\Ne)\xi_\phi(\Ne^\prime)}=\braket{\xi_{\psi_r}(\Ne)\xi_{\psi_r}(\Ne^\prime)}=\delta(\Ne-\Ne^\prime) \qc \braket{\xi_\phi(\Ne)\xi_{\psi_r}(\Ne^\prime)}=0,
}
The stochastic noise for the inflaton $\phi$ is usually negligible for 
the waterfall dynamics, whereas that for the waterfall fields is important around the waterfall phase transition. 
We denote the e-folding number at the time of the waterfall phase transition as $\Nc$. 
Neglecting the stochastic noise for $\phi$, 
it is solved as
\bae{
    \phi \simeq \phi_\uc - \frac{\Mpl^2 (\Ne-\Nc)}{ \mu_1}, 
}
for $\Ne \approx \Nc$. 
\eq{eq:stochastic_r} 
derives the evolution equation 
for $\braket{\psi_r^2}$ as 
\bae{
    \dv{\braket{\psi_r^2}}{\Ne} 
    \simeq\pqty{\frac{4}{\Pi}}^2 (\Ne-\Nc) \braket{\psi_r^2} + \D \frac{V}{12\pi^2\Mpl^2}, 
}
where we assume $\braket{\psi_r^2} \ll M^2$ and define 
\bae{
    \Pi \equiv \frac{M \sqrt{\mu_1 \phi_c}}{\Mpl^2}. 
    \label{Pi}
}
With an approximation $V\simeq\Lambda^4$, it can be solved 
as 
\bae{
    \label{eq:r}
    \braket{\psi_r^2}(\Ne) 
	\simeq\psi_{r,\uc}^2 \bqty{1 + \Erf\pqty{\frac{2\sqrt{2} (\Ne-\Nc)}{\Pi}} } \exp \bqty{\frac{8 (\Ne-\Nc)^2}{\Pi^2}}, 
}
where we adopt the asymptotic condition $\braket{\psi_r^2}\to0$ for $\Ne \ll \Nc$. 
Here, $\Erf(x) \equiv \frac{2}{\sqrt{\pi}} \int_0^x \ee^{-t^2} \dd{t}$ is the error function. 
$\psi_{r,\uc}$ is defined by
\bae{
    \psi_{r,\uc}^2 = \frac{\D \Lambda^4 \Pi}{48 \sqrt{2 \pi^3} \Mpl^2 }, 
    \label{eq:r0}
}
hence represents the amplitude of $\braket{\psi_r^2}$ at the phase transition ($\Ne=\Nc$).
This is enhanced by a factor of $\D$ for multiple waterfall fields. 

The solution \eq{eq:r} approaches to $2 \psi_{r,\uc}^2 \exp[8 (\Ne-\Nc)^2/\Pi^2]$ for $\Ne \gg \Pi/(2\sqrt{2}) \equiv \Ne_{\rm cl}$. 
The time scale $\Ne_{\rm cl}$ represents the time after which the classical dynamics (namely the effect of the first term in the right-hand side of \eq{eq:stochastic_r}) begins to dominate. 
We will observe later that $\Pi$ should be $10\text{--}40$, so that $\Ne_{\rm c} =7\text{--}28$ for the case we are interested in.
The stochastic effect is therefore important even well after the waterfall phase transition. 
As will be observed below, large curvature perturbations are generated during this regime, so that the numerical simulation for the \ac{FP} equation is necessary to calculate the spectrum of curvature perturbations. This will be done in Sec.~\ref{sec:simulation}. 
In this subsection, we nevertheless 
derive an analytic rough estimation on the curvature perturbation by omitting the stochastic effect and solving the classical equation of motion in the waterfall phase with the initial condition of $\psi_r = \psi_{r,\uc}$ at $\Ne = \Nc$, following Refs.~\cite{Clesse:2015wea,Kawasaki:2015ppx,Tada:2023pue}.
We will see that it indeed
provides some information of the spectrum of the curvature perturbation. 

Here we quote some results presented in Ref.~\cite{Tada:2023pue} 
with some corrections by a factor of $\D$ 
including the leading-order correction from the quadratic potential for the inflaton. 
The e-folding number from the time of the waterfall phase transition $\Nc$ to the end of inflation $\Nend$ is calculated as 
\bae{
    \Ne_{\rm PT} 
    \equiv \Ne_{\rm end} - \Nc \simeq
    \Pi \lmk \frac{\sqrt{\chi_2}}{2} + \frac{c}{4 \sqrt{\chi_2}} \rmk - \Pi^2 \frac{\Mpl^2}{12 \mu_2^2} \lmk \chi_2 + c \rmk, 
    \label{eq:NPT0}
}
where $c$ ($<1$) and $\chi_2$ are given by 
\bae{
    &c \equiv
    \int_{\chi_2}^{\chi_{\rm end}} \frac{\dd{\chi}}{\sqrt{1 + M^2 \bqty{\ee^{2 (\chi - \chi_2)} - 1} / (8 \mu_1 \phi_\uc \infl_2^2) 
    }},
    \label{eq:c}
    \\
    &\chi_2 = \frac{1}{2} \ln \lmk \frac{12 \sqrt{2 \pi^3} \Mpl^6 \Pi}{\D \Lambda^4 \mu_1^2} \rmk. 
    \label{chi1}
}
The $\delta N$ approach gives the spectrum of curvature perturbations 
as\footnote{
The factor of $\chi_2 / \Delta N_1^2$ in the exponent should be replaced by $4/ \Pi^2$ in the analytic formula if one consistently perform the perturbation with respect to $(\Mpl/\mu_2)^2$. However, we instead adopt \eq{eq:width0} because it fits better for our numerical results. One can regard \eq{eq:width0} as a fitting fomula rather than the analytic one. 
}
\bae{
    {\cal P}_\curv (k)
    = {\cal P}_\curv^{(\rm peak)} \exp 
    \lkk - 2 \chi_2 \frac{ (\Ne_\mathrm{PT}-\calNhat_k)^2 + (2 \Mpl^2/(3 \mu_2^2)) \lmk \Ne_\mathrm{PT}-\calNhat_k \rmk^3}{ \Delta \Ne_1^{2}} \rkk,  
    \label{eq:width0}
}
where the backward e-folds $\calNhat_k$ is defined by the time when $k=aH$ and $\Delta \Ne_1$ is the solution to the following equation: 
\bae{
    \chi_2 = \frac{4}{\Pi^2} \Delta \Ne_1^2 
    + \frac{8\Mpl^2}{3\mu_2^2\Pi^2} \Delta \Ne_1^3.
    \label{eq:N1chi}
}
The peak amplitude reads 
\bae{
    {\cal P}_\curv^{(\rm peak)}
    &\simeq
    \frac{\Pi}{2 \sqrt{2\pi}  \chi_2 \D} 
    \lmk 
    1 -   \frac{\Mpl^2}{3 \mu_2^2} \Pi \sqrt{\chi_2}
    \rmk
    \label{eq:Ppeak-Pi0}
    \\
    &\simeq 0.013 \, \Ne_{\rm PT} \frac{1}{\D} \lmk \frac{\chi_2}{10} \rmk^{-3/2} \lmk 
    1 -   \frac{\Mpl^2}{3 \mu_2^2} 
    \Ne_{\rm PT}
    \rmk,
    \label{eq:Pzeta0}
}
where we neglect $\mathcal{O}(c/\chi_2)$ terms in the second line.

We will see in Sec.~\ref{sec:parameters} that the parameter $\chi_2$ is $\sim 10$ almost constantly for the parameters of interest and also $\mu_2$ is fixed to $\simeq10\Mpl$ by the \ac{CMB} observation. 
Therefore, the peak amplitude $\calP_\curv^{\mathrm{(peak)}}$ is uniquely determined by the corresponding peak wavelength (through $\Ne_\mathrm{PT}$) except for the suppression due to $\D$. 
If $\D$ is equal to unity or two, 
the peak amplitude is too large for a desirable amount of \acp{PBH} with astrophysical scales~\cite{Clesse:2015wea,Kawasaki:2015ppx}. 
This challenge can be evaded when we consider multi-waterfall fields because of the factor of $1/\D$ in \eq{eq:Pzeta0}.

As we commented above, the stochastic effect is not negligible for several e-foldings after the waterfall phase transition. 
Nevertheless, the above analytic form works reasonably well as we confirm with 
the detailed numerical calculations  
in the stochastic formalism in the next subsection. We will use the analytic formula~\eqref{eq:width0} as a fitting function of the numerical result, regarding $\Ne_\mathrm{PT}$ and $\calP_\curv^\mathrm{(peak)}$ as fitting parameters rather than giving them by Eqs.~\eqref{eq:NPT0} and \eqref{eq:Ppeak-Pi0}.

\subsection{Numerical approach}
\label{sec:simulation}

\subsubsection[Numerical implementation of the stochastic-$\delta N$ formalism]{\boldmath Numerical implementation of the stochastic-$\delta N$ formalism}

In the full stochastic approach to inflation, the power spectrum of the curvature perturbation can be calculated in the so-called \emph{stochastic-$\delta N$} formalism~\cite{Fujita:2013cna,Fujita:2014tja,Vennin:2015hra,Ando:2020fjm,Tada:2021zzj}.
The essential idea is that the probability distribution of the stochastic backward e-folds $\calNhat$ is nothing but that of the curvature perturbation thanks to the $\delta N$ formalism~\cite{Starobinsky:1985ibc,Sasaki:1995aw,Wands:2000dp,Lyth:2004gb}.
The power spectrum is obtained by extracting the variance of $\calNhat$ corresponding to the scale of interest.
Specifically, it is given by the field-space integration (see Eq.~(3.19) of Ref.~\cite{Tada:2021zzj})
\bae{\label{eq: power in stoc-dN}
    \calP_\curv(k)=-\int_\Omega\dd{\bm{\phi}_*}\pdv{P_\bw(\bm{\phi}_*\mid\calNhat_k)}{\calNhat_k}\braket{\delta\calNhat^2(\bm{\phi}_0\to\bm{\phi}_*)},
}
weighted by the backward probability distribution $P_\bw(\bm{\phi}\mid\calNhat)$, the probability such that the inflatons take certain field values $\bm{\phi}=\Bqty{\phi,\psi_r}$ at the time $\calNhat$ e-folds before the end of inflation.
$\Omega$ denotes the field-space region where the slow-roll inflation can be realised. $\bm{\phi}_0$ is certain initial field values of the inflatons and $\delta\calNhat(\bm{\phi}_0\to\bm{\phi}_*)$ is a fluctuation in the e-folds $\calNhat(\bm{\phi}_0\to\bm{\phi}_*)$ from $\bm{\phi}_0$ to $\bm{\phi}_*$, i.e., $\delta\calNhat(\bm{\phi}_0\to\bm{\phi}_*)\equiv\calNhat(\bm{\phi}_0\to\bm{\phi}_*)+\braket{\calNhat(\bm{\phi}_*)}-\braket{\calNhat(\bm{\phi}_0)}$, where $\calNhat(\bm{\phi})$ is a first passage time from $\bm{\phi}$ to the end of inflation.
Note that this formula requires the leading-order slow-roll approximation, $\calNhat_k\simeq-\ln(k/k_\uf)$, where $k_\uf$ is the comoving Hubble scale $aH$ at the end of inflation.

While the backward probability $P_\bw$ is formulated by a combination of the probability density function of $\bm{\phi}$ (the solution of the \ac{FP} equation) and that of $\calNhat$ (the solution of the adjoint \ac{FP} equation) in a rigorous way (see Ref.~\cite{Ando:2020fjm}), it is easy to sample the solution of the Langevin equation in a practical, numerical simulation. There, the backward probability is approximated as
\bae{
    P_\bw(\bm{\phi}\mid\calN)\approx\frac{1}{S}\sum_{i=1}^S\delta\pqty{\bm{\phi}-\bm{\phi}_i(\calNhat_i(\bm{\phi}_0)-\calNhat)},
}
where $S$ is the sampling number, $\bm{\phi}_i(N)$ is the $i$th solution of the Langevin equation, and $\calNhat_i$ is the first passage time from the initial field value $\bm{\phi}_0$ for that solution.
Its derivative can be further approximated by the finite difference:
\bae{
    \pdv{P_\bw(\bm{\phi}\mid\calNhat)}{\calNhat}\approx\frac{P_\bw(\bm{\phi}\mid\calNhat+\Delta N/2)-P_\bw(\bm{\phi}\mid\calNhat-\Delta N/2)}{\Delta N},
}
with a certain small step $\Delta N$.
Substituting these approximations back into the formula~\eqref{eq: power in stoc-dN}, the power spectrum is approximated as
\bae{
    \calP_\curv(k)\approx\frac{1}{S\times\Delta N}\sum_{i=1}^S\left[\Braket{\delta\calNhat^2\pqty{\bm{\phi}_0\to\bm{\phi}_i^+(k)}}-\Braket{\delta\calNhat^2\pqty{\bm{\phi}_0\to\bm{\phi}_i^-(k)}}\right],
}
where $\bm{\phi}_i^\pm(k)=\bm{\phi}_i\pqty{\calNhat_i(\bm{\phi}_0)-\calN_k\pm\frac{\Delta N}{2}}$.

Note that the average $\Braket{\delta\calNhat^2\pqty{\bm{\phi}_0\to\bm{\phi}_i^\pm(k)}}$ is defined only for sample paths passing the point $\bm{\phi}_i^\pm(k)$, which is a non-trivial condition in a two or more than two fields case. We further try to approximate it by the average without a condition.
Let us first define the total perturbation for paths passing $\bm{\phi}_i^\pm(k)$ by
\bae{
    \eval{\delta\calNhat(\bm{\phi}_0)}_{\bm{\phi}_i^\pm(k)}&=\calNhat\qty(\bm{\phi}_0\to\bm{\phi}_i^\pm(k))+\calNhat(\bm{\phi}_i^\pm(k))-\Braket{\calNhat(\bm{\phi}_0)} \nonumber \\
    &=\calNhat\qty(\bm{\phi}_0\to\bm{\phi}_i^\pm(k))+\Braket{\calNhat\qty(\bm{\phi}_i^\pm(k))}-\Braket{\calNhat(\bm{\phi}_0)}+\calNhat\qty(\bm{\phi}_i^\pm(k))-\Braket{\calNhat\qty(\bm{\phi}_i^\pm(k))} \nonumber \\
    &=\delta\calNhat\qty(\bm{\phi}_0\to\bm{\phi}_i^\pm(k))+\delta\calNhat\qty(\bm{\phi}_i^\pm(k)).
}
Mean squares of both sides read
\bae{
    \Braket{\delta\calNhat^2(\bm{\phi}_0)}_{\bm{\phi}_i^\pm(k)}&=\Braket{\Bigl(\delta\calNhat\qty(\bm{\phi}_0\to\bm{\phi}_i^\pm(k))+\delta\calNhat\qty(\bm{\phi}_i^\pm(k))\Bigr)^2} \nonumber \\
    &=\Braket{\delta\calNhat^2\qty(\bm{\phi}_0\to\bm{\phi}_i^\pm(k))}+\Braket{\delta\calNhat^2\qty(\bm{\phi}_i^\pm(k))},
}
where we used the Markovian property of the Langevin equation for the last line.
We then expect that the difference between $\Braket{\delta\calNhat^2(\bm{\phi}_0)}_{\bm{\phi}_i^+(k)}$ and $\Braket{\delta\calNhat^2(\bm{\phi}_0)}_{\bm{\phi}_i^-(k)}$ will be suppressed as $\calO(\Delta N^2)$. This would be statistically justified because for typical points $\bm{\phi}_i^\pm(k)$, $\Braket{\delta\calNhat^2(\bm{\phi}_0)}_{\bm{\phi}_i^\pm(k)}$ asyptotes to a function only of $\bm{\phi}_0$ and becomes stationary against the change between $\bm{\phi}_i^+(k)$ and $\bm{\phi}_i^-$.
Therefore, one finds
\bae{\label{eq: calP in StocDeltaN}
    \calP_\curv(k)\approx\frac{1}{S\times\Delta N}\sum_{i=1}^S\bqty{\Braket{\delta\calNhat^2(\bm{\phi}_i^-(k))}-\Braket{\delta\calNhat^2(\bm{\phi}_i^+(k))}},
}
at the leading order in $\Delta N$.

The variances on the right-hand side are found as a solution of the adjoint \ac{FP} equation. In fact, recalling that the average $\Braket{f\qty(\calNhat(\bm{\phi}))}$ of an arbitrary function $f\qty(\calNhat(\bm{\phi}))$ is defined by the probability density $P_\FPT$ as
\bae{
    \Braket{f\qty(\calNhat(\bm{\phi}))}=\int\dd{\calNhat}f(\calNhat)P_\FPT(\calNhat\mid\bm{\phi}),
}
the adjoint \ac{FP} equation
\bae{
    \partial_\calNhat P_\FPT(\calN\mid\bm{\phi})&=\calL_\FP^\dagger\cdot P_\FPT(\calN\mid\bm{\phi}) \nonumber \\
    &\bmbe{\equiv\left[\pqty{-\Mpl^2\frac{V_\phi}{V}}\partial_\phi+\pqty{-\Mpl^2\frac{V_{\psi_r}}{V}+\frac{1}{2}\frac{V}{12\pi^2\Mpl^2}\frac{\D-1}{\psi_r}}\partial_{\psi_r} \right. \\
    \left.+\frac{1}{2}\frac{V}{12\pi^2\Mpl^2}\pqty{\partial_\phi^2+\partial_{\psi_r}^2}\right]P_\FPT(\calN\mid\bm{\phi}),}
}
with the boundary condition $P_\FPT(\calN\mid\bm{\phi})=\delta(\calNhat)$ on the end-of-inflation surface $\partial\Omega$ leads to the following recursive partial differential equations:
\bae{\label{eq: recursive PDE}
    \bce{
        \dps
        \calL_\FP^\dagger\cdot\braket{\calNhat(\bm{\phi})}=-1, \\
        \dps
        \calL_\FP^\dagger\cdot\braket{\delta\calNhat^2(\bm{\phi})}=-\frac{V}{12\pi^2\Mpl^2}\bqty{\Bigl(\partial_\phi\braket{\calNhat(\bm{\phi})}\Bigr)^2+\Bigl(\partial_{\psi_r}\braket{\calNhat(\bm{\phi})}\Bigr)^2},
    }
}
with the boundary condition $\eval{\braket{\calNhat(\bm{\phi})}}_{\bm{\phi}\in\partial\Omega}=\eval{\braket{\delta\calNhat^2(\bm{\phi})}}_{\bm{\phi}\in\partial\Omega}=0$.
In this paper, we numerically solve these partial differential equations in the Jacobi method with use of the \CC package, \textsc{StocDeltaN}~\cite{StocDeltaN} (see also the author's GitHub page~\cite{github}).

\subsubsection{Results}

We show the numerical results of the stochastic-$\delta N$ formalism in this subsubsection.
We fixed $M$ and $\phi_\uc$ by $M=\phi_\uc/\sqrt{2}=10^{16}\,\si{GeV}$ while $\mu_1$ is determined by $\Pi$ parameter through its definition~\eqref{Pi}. $\Lambda$ and $\mu_2$ are fixed by Eqs.~\eqref{eq: Lambda} and \eqref{ns} due to the \ac{CMB} constraint.
In the $\delta N$ formalism, the end surface should be given by a uniform-density slice. We first solve the classical slow-roll equations of motion (Eq.~\eqref{eq:stochastic_r} without the noise terms) from the initial condition $\phi=\phi_\uc$ and $\psi=\psi_{r,\uc}$ until $\eta_{\psi\psi}\equiv\Mpl^2\frac{V_{\psi_r\psi_r}}{V}$ reaches $-2$ (for sufficient violation of the slow-roll condition) for the first time and define the end-of-inflation energy density for the stochastc-$\delta N$ scheme by the potential value at that time.

\begin{figure}
    \centering
    \begin{tabular}{ll}
        \begin{minipage}{0.48\hsize}
            \includegraphics[width=0.95\hsize]{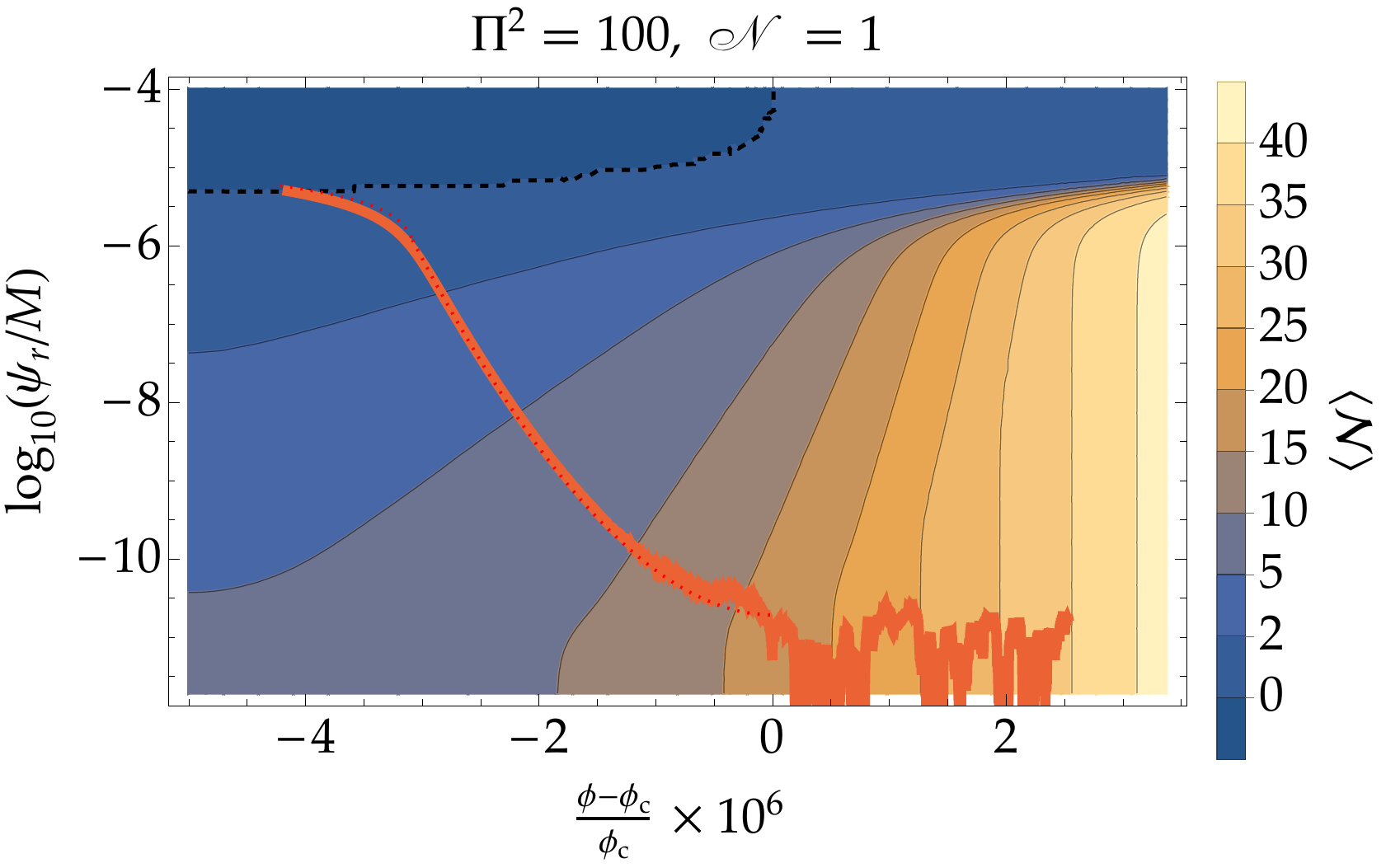}
        \end{minipage} & \hspace{-10pt}
        \begin{minipage}{0.48\hsize}
            \includegraphics[width=0.95\hsize]{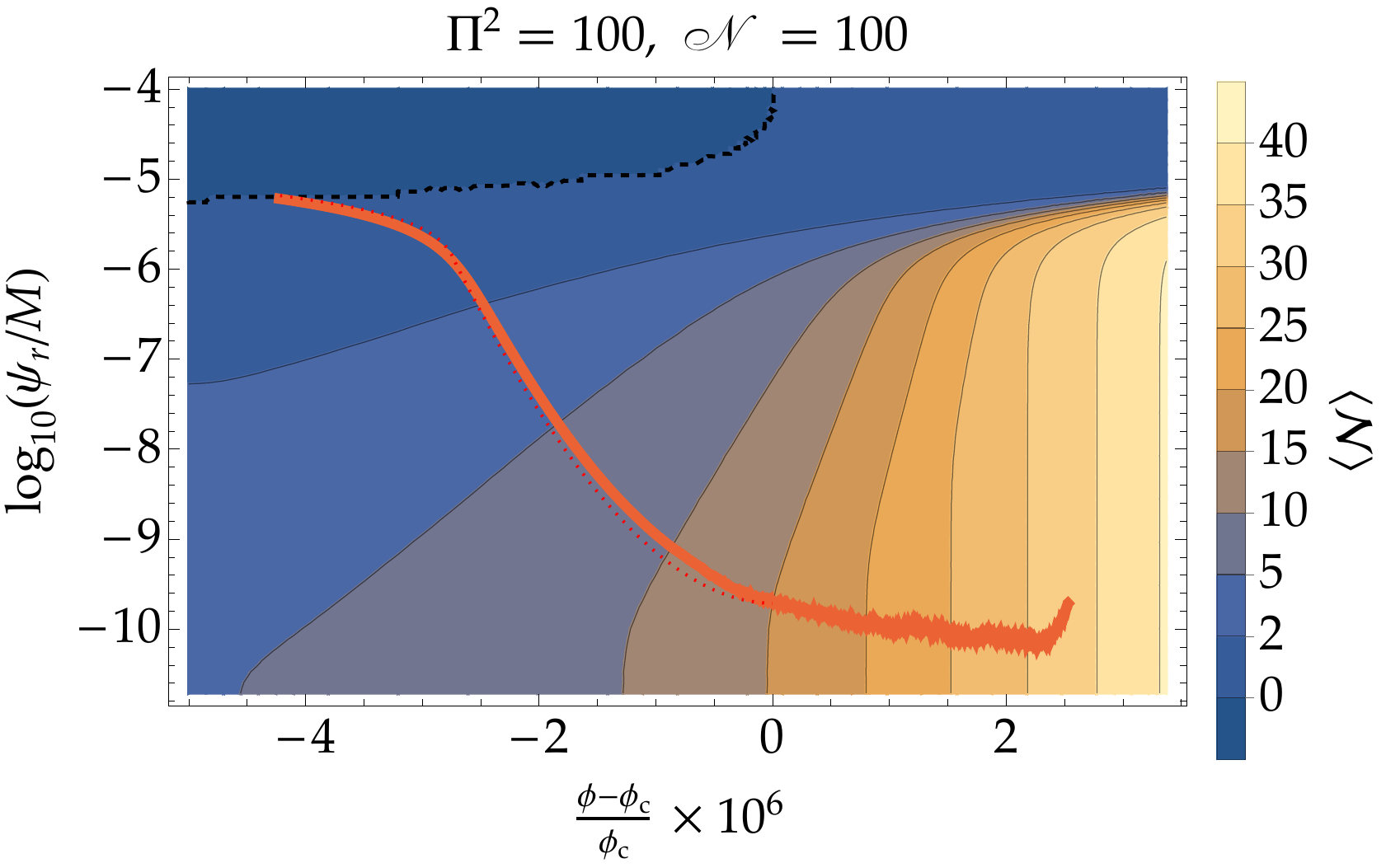}
        \end{minipage} \\
        \begin{minipage}{0.5\hsize}
            \includegraphics[width=0.95\hsize]{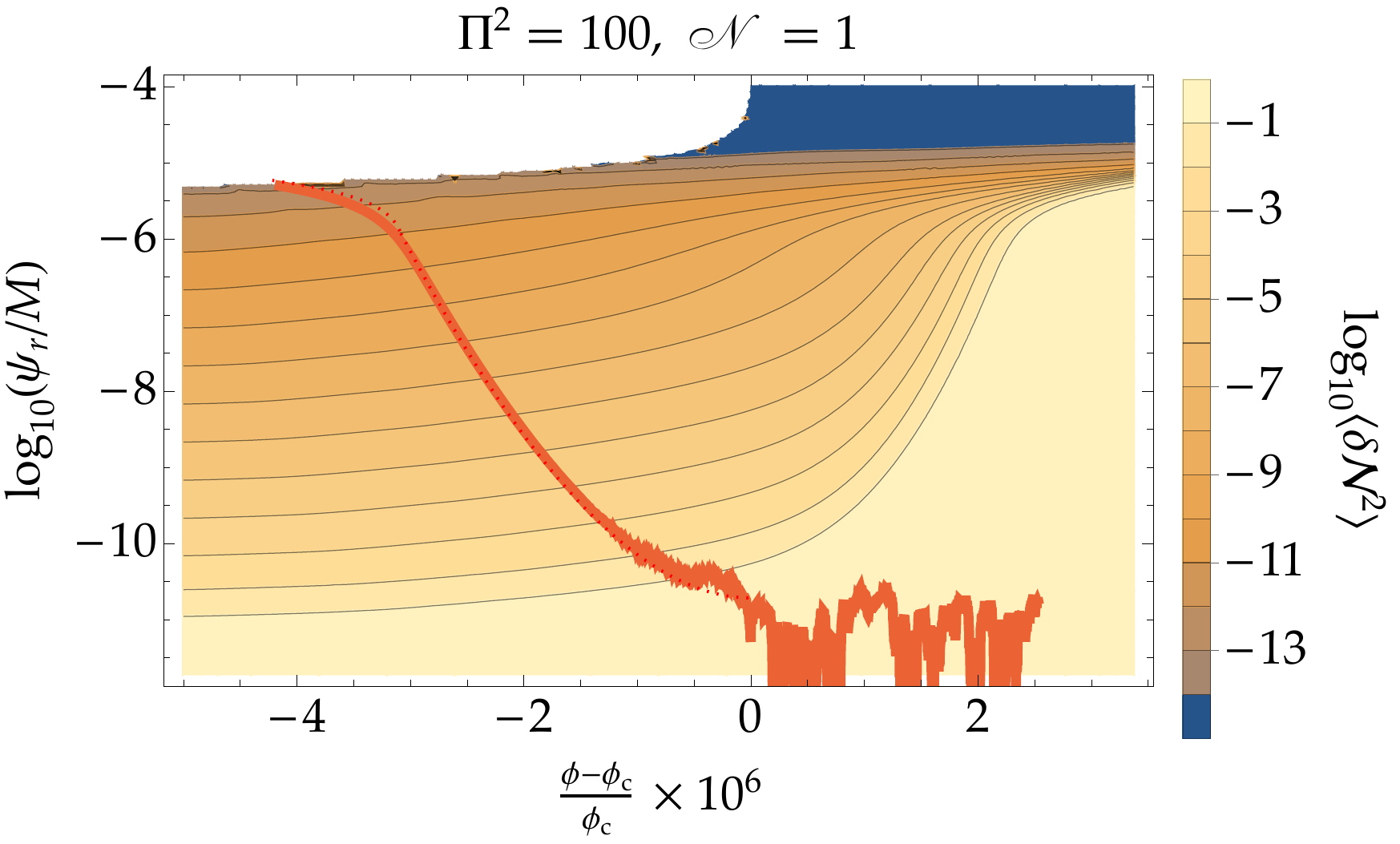}
        \end{minipage} & \hspace{-10pt}
        \begin{minipage}{0.5\hsize}
            \includegraphics[width=0.95\hsize]{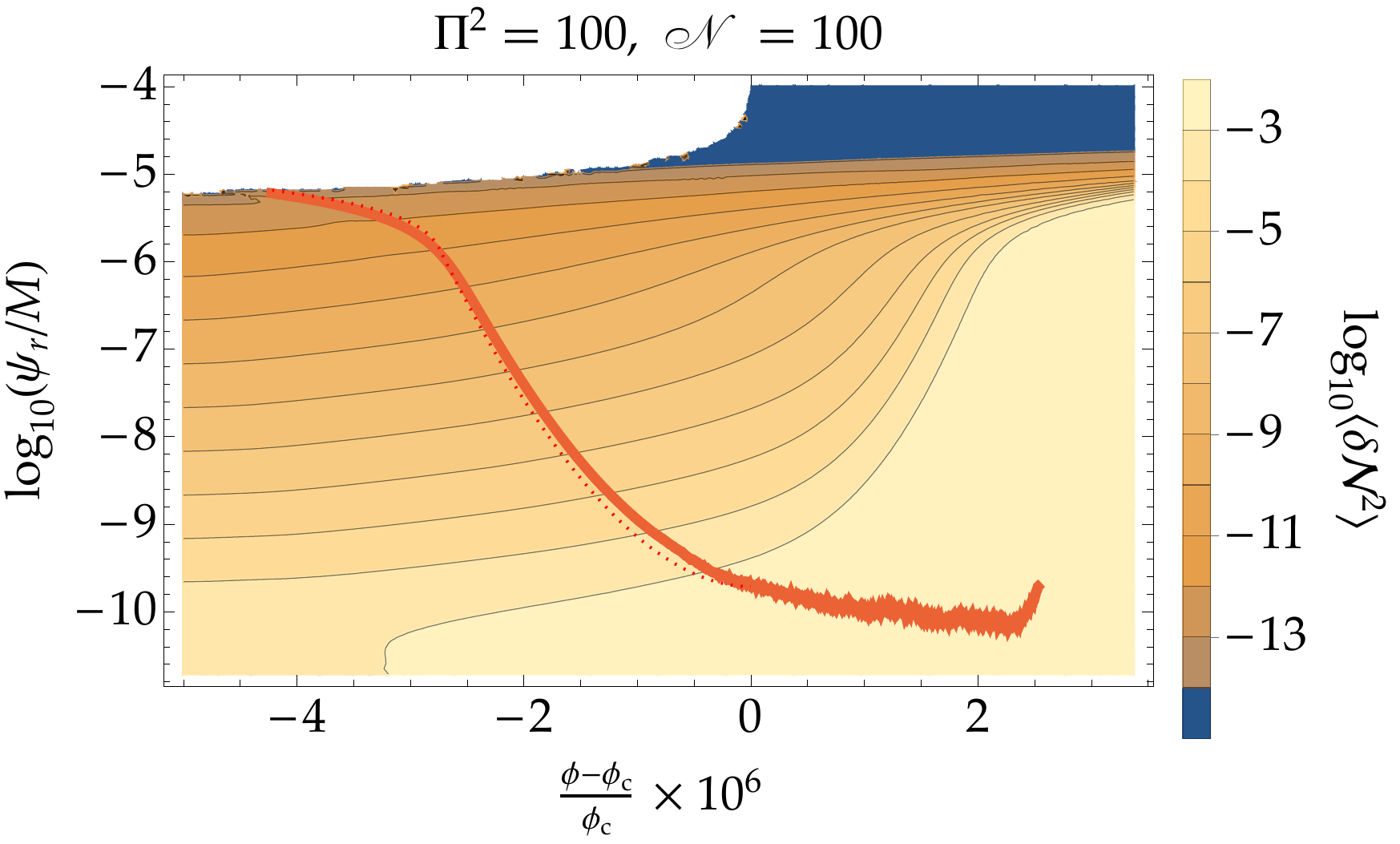}
        \end{minipage}        
    \end{tabular}
    \caption{Contours of $\braket{\calNhat(\phi,\psi_r)}$ for $\D=1$ (top left) and $100$ (top right)
    and $\braket{\delta\calNhat^2(\phi,\psi_r)}$ for $\D=1$ (bottom left) and $100$ (bottom right) with $\Pi^2= 100$. 
    Red thick lines are sample solutions of the Langevin equations while red dotted ones are classical solutions without the noise terms. 
    Black dotted lines indicate the end-of-inflation surface.
    }
    \label{fig: N}
\end{figure}

Fig.~\ref{fig: N} shows the contours of $\braket{\calNhat(\phi,\psi_r)}$ and $\braket{\delta\calNhat^2(\phi,\psi_r)}$ for $\Pi^2=100$ and $\D=1$ or $100$ as solutions of the recursive partial differential equations~\eqref{eq: recursive PDE} obtained by the \textsc{StocDeltaN} package. One finds that inflatons' sample trajectories represented by red lines pass the larger-$\psi_r$ region for $\D=100$ than $\D=1$ due to the noise-induced centrifugal force. The interesting feature is that though $\braket{\calNhat}$ is not so sensitive to the number of waterfall fields $\D$, the perturbation variance $\braket{\delta\calNhat^2}$ indeed decreases in the large $\D$ case. Therefore, one can expect that a large $\D$ can reduce $\calP_\curv$, keeping its peak scale large enough.

\begin{figure}
    \centering
    \begin{tabular}{c}
        \begin{minipage}{0.49\hsize}
            \centering
            \includegraphics[width=0.95\hsize]{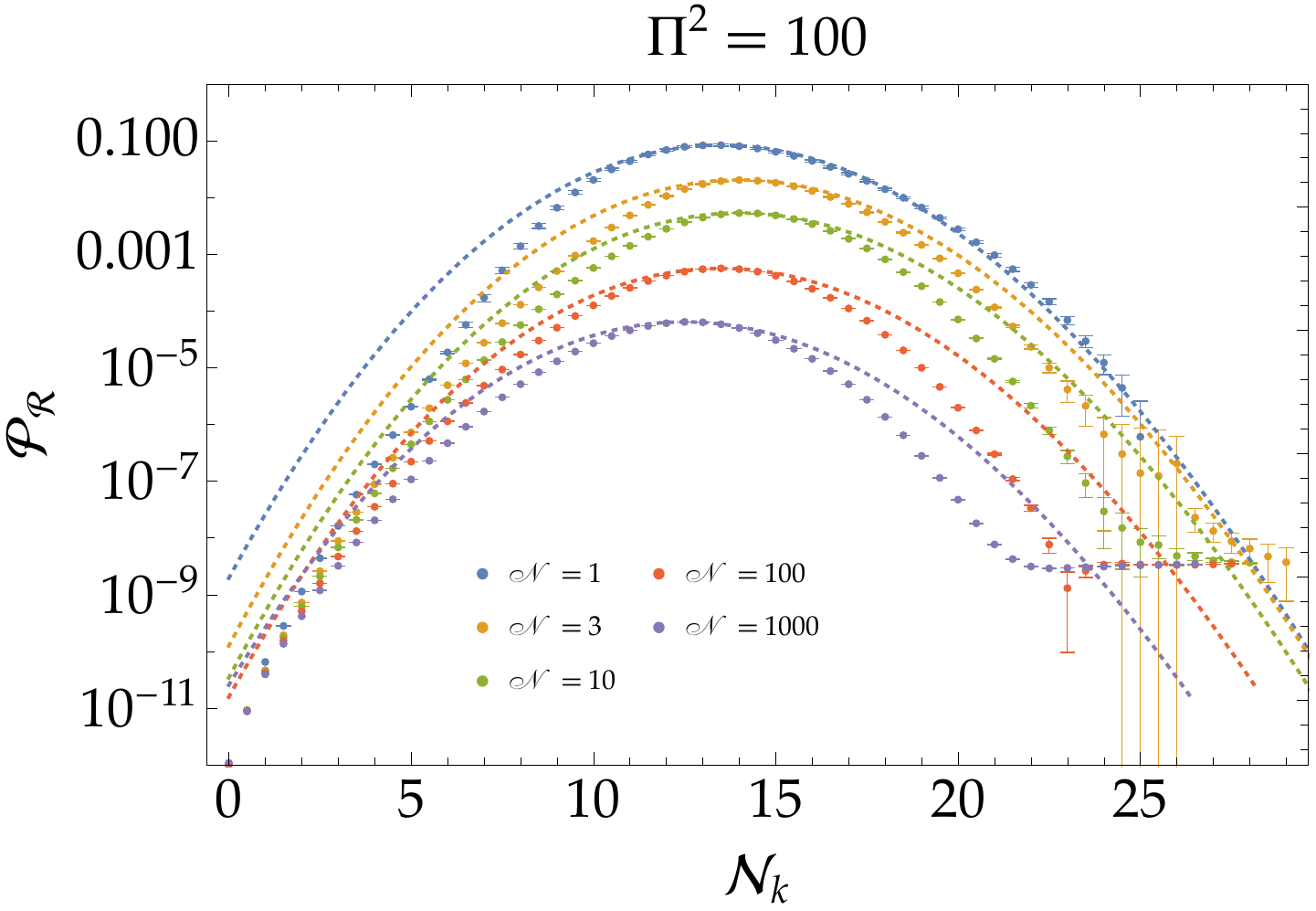}
        \end{minipage}
        \begin{minipage}{0.49\hsize}
            \centering
            \includegraphics[width=0.95\hsize]{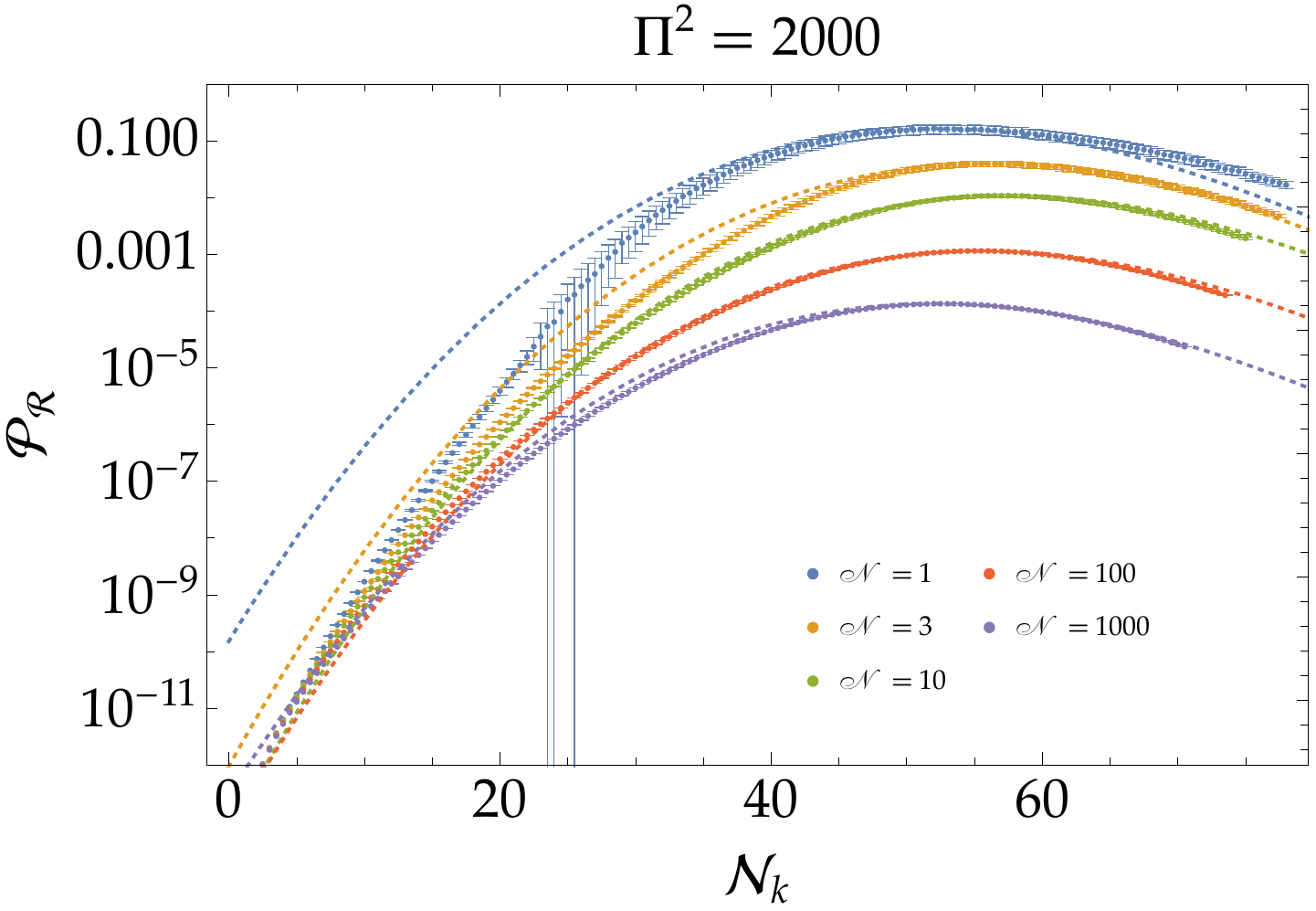}
        \end{minipage} 
    \end{tabular}
    \caption{Power spectra of curvature perturbations obtained in the stochastic-$\delta N$ formalism for $\D=1$ (blue dots), $3$ (orange dots) $10$ (green dots), $100$ (red dots), and $1000$ (purple dots) for $\Pi^2=100$ (left) 
    and $2000$ (right). The sampling number is $S=10^5$ and the e-folds step is $\Delta\Ne=0.5$. Each error bar represents the corresponding standard error. 
    Coloured dotted lines show the corresponding fitting formula~\eqref{eq:width0} with the fitting parameters $\Ne_\mathrm{PT}$ and $\calP_\curv^\mathrm{(peak)}$ chosen by the numerically obtained peak wavenumber and peak amplitude.}
    \label{fig: calP}
\end{figure}

Fig.~\ref{fig: calP} shows the resultant power spectra of curvature perturbations for several $\Pi^2$ and $\D$. We adopt the sampling number $S=10^5$ and the e-folds step $\Delta\Ne=0.5$. Each error bar represents the propagated error in the finite difference equation~\eqref{eq: calP in StocDeltaN} from the estimated standard error in the sampling averages $\frac{1}{S}\sum_{i=1}^S\braket{\delta\calNhat^2(\bm{\phi}_i^\pm(k))}$.
We also show the fitting formula~\eqref{eq:width0} by the dotted lines with the corresponding colours. The fitting parameters $\Ne_\mathrm{PT}$ and $\calP_\curv^\mathrm{(peak)}$ are chosen by the numerically obtained peak wavenumber and peak amplitude.
One finds that the fitting formula reasonably works well, particularly for a large $\Pi^2$ and a large $\D$ where the slow-roll approximation and the classical approximation (to neglect the noise terms) are better respectively.

\begin{figure}
    \centering
    \begin{tabular}{c}
        \begin{minipage}{0.49\hsize}
            \centering
            \includegraphics[width=0.95\hsize]{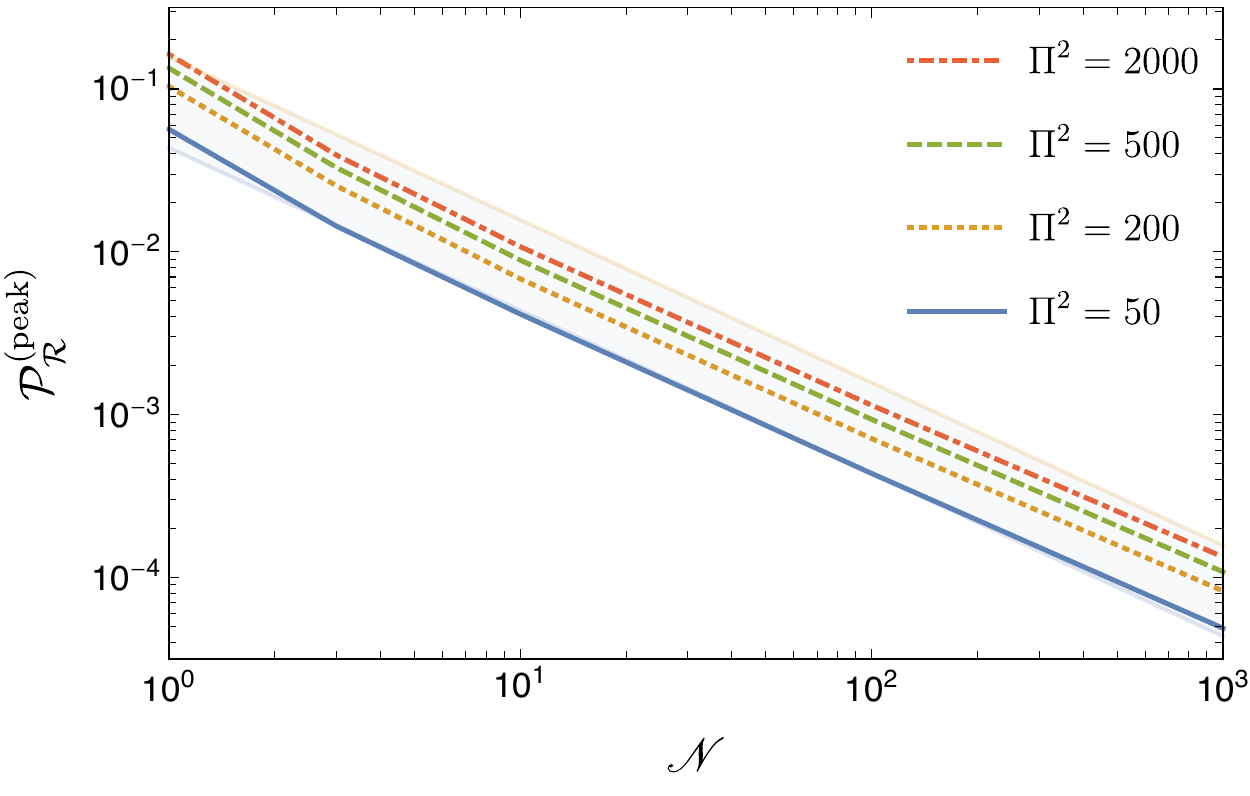}
        \end{minipage}
        \begin{minipage}{0.49\hsize}
            \centering
            \includegraphics[width=0.95\hsize]{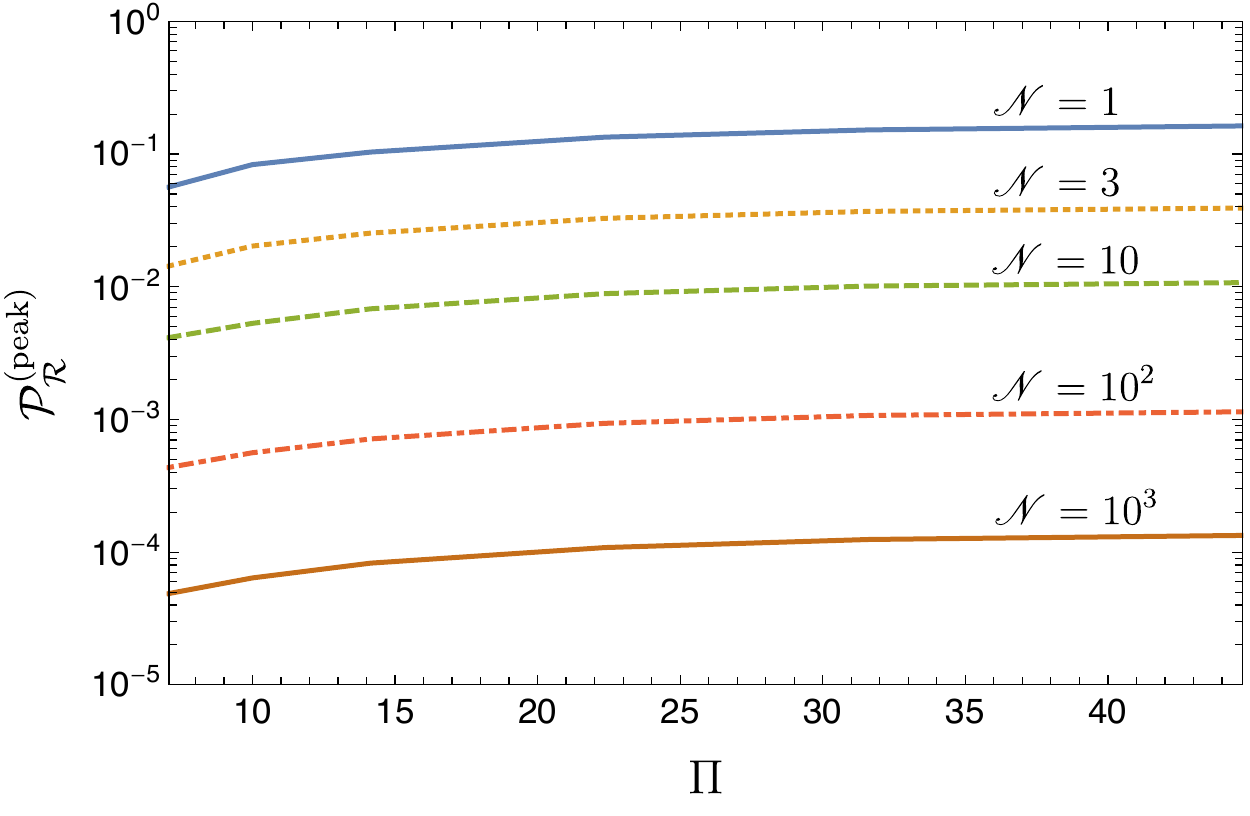}
        \end{minipage} \\
        \begin{minipage}{0.49\hsize}
            \centering
            \includegraphics[width=0.95\hsize]{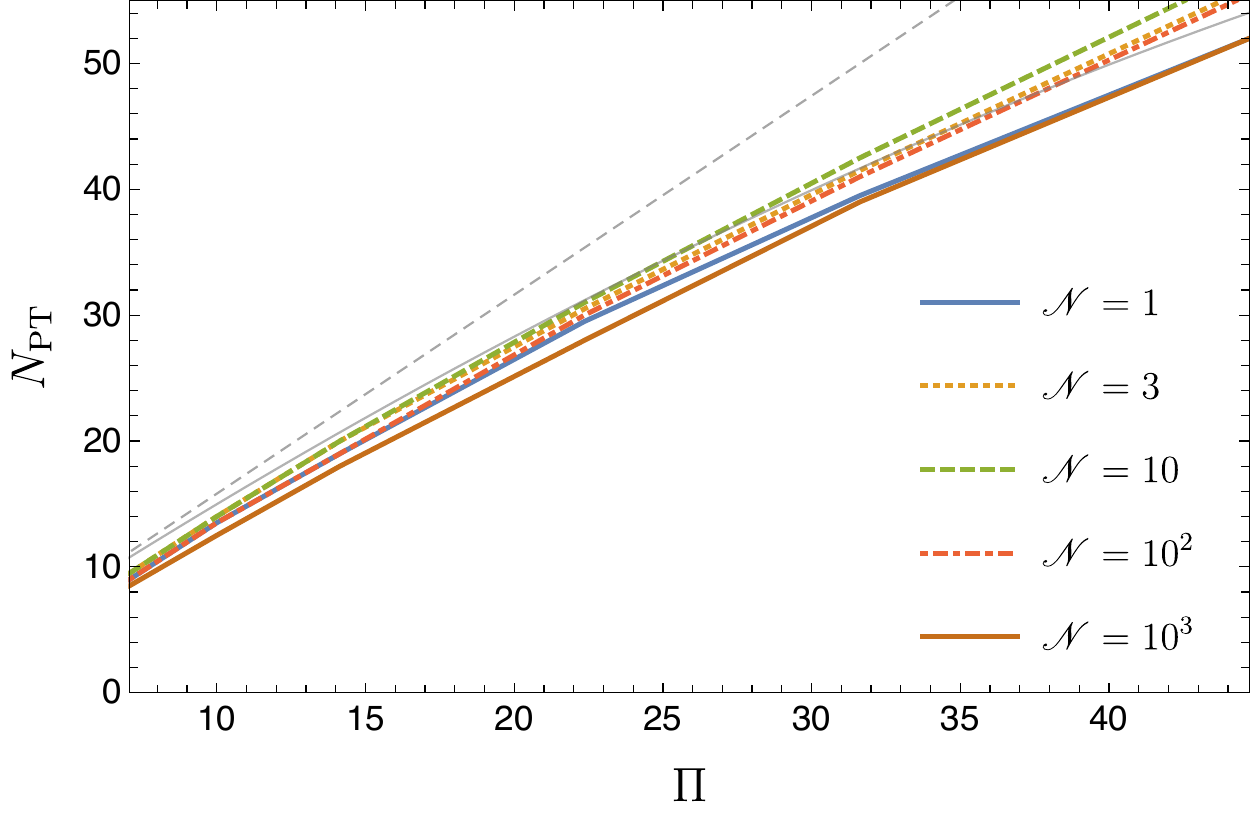}
        \end{minipage}
    \end{tabular}
    \caption{The numerical result for the fitting parameters $\calP_\curv^\mathrm{(peak)}$ (top) and $\Ne_{\rm PT}$ (bottom) as functions of $\Pi$ and $\D$. 
    On the top-left panel, the shaded region represents the analytical result \eq{eq:Ppeak-Pi0} multiplied by a factor of $1/3$ for $\Pi^2 \in (50, 2000)$.
    The thin dashed (solid) curve 
    on the bottom panel represents the analytic 
    formula \eq{eq:NPT0} in the leading $\sim\Pi$ (next-to-leading $\sim\Pi^2$) order approximations with $c=0$. 
    }
    \label{fig: NPT and calPpeak}
\end{figure}

Table~\ref{tab:1} shows the numerical results of the fitting parameters $\Ne_\mathrm{PT}$ and $\calP_\curv^\mathrm{(peak)}$ for $\Pi^2 = 50$, $100$, $200$, $500$, $1000$, and $2000$ with $\D = 1$, $3$, $10$, $100$, and $1000$. 
Those results are also shown in Fig.~\ref{fig: NPT and calPpeak}. 
The shaded region on the top-left panel represents the analytical result \eq{eq:Ppeak-Pi0} multiplied by a factor of $1/3$ for $\Pi^2 \in (50, 2000)$. 
One finds that the analytic formula reasonably works again. 
Furthermore, the behaviour $\calP_\curv^\mathrm{(peak)}\propto1/\D$ is confirmed by the numerical result. 
The thin dashed (solid) curves on the bottom panel represent the analytic results \eq{eq:NPT0} in the leading $\sim\Pi$ (next-to-leading $\sim\Pi^2$) order approximations with $c = 0$.
Though the leading order approximation is not enough, the next-to-leading order approximation is well consistent with the numerical result. One can also see that $\Ne_\mathrm{PT}$ is almost independent of $\D$ and hence the degeneracy between $\Ne_\mathrm{PT}$ and $\calP_\curv^\mathrm{(peak)}$ can be solved by introducing $\D$.

\begin{table*}
\begin{center}
\begin{tabular}{cccc}
\hline
   $\Pi^2$ & $\D$ & $\Ne_\mathrm{PT}$ & $\calP_\curv^\mathrm{(peak)}$ \\[.2em]
	\hline \hline
 50 & 1 & 9 & 0.056 \\ 
 50 & 3 & 9.5 & 0.014 \\ 
 50 & 10 & 9.5 & 0.0041 \\ 
 50 & 100 & 9 & 0.00044 \\ 
 50 & 1000 & 8.5 & 0.000049 \\ 
	\hline
 100 & 1 & 13.5 & 0.083 \\ 
 100 & 3 & 14 & 0.020 \\ 
 100 & 10 & 14 & 0.0053 \\ 
 100 & 100 & 13.5 & 0.00056 \\ 
 100 & 1000 & 12.5 & 0.000064 \\ 
	\hline
 200 & 1 & 19 & 0.10 \\ 
 200 & 3 & 20 & 0.025 \\ 
 200 & 10 & 20 & 0.0068 \\ 
 200 & 100 & 19 & 0.00071 \\ 
 200 & 1000 & 18 & 0.000083 \\ 
\hline 
\end{tabular}
\qquad
\begin{tabular}{cccc}
\hline
   $\Pi^2$ & $\D$ & $\Ne_\mathrm{PT}$ & $\calP_\curv^\mathrm{(peak)}$ \\[.2em]
	\hline \hline
 500 & 1 & 29.5 & 0.13 \\ 
 500 & 3 & 30.5 & 0.033 \\ 
 500 & 10 & 31 & 0.0089 \\ 
 500 & 100 & 30 & 0.00094 \\ 
 500 & 1000 & 28 & 0.00011 \\ 
	\hline
 1000 & 1 & 39.5 & 0.15 \\ 
 1000 & 3 & 41.5 & 0.037 \\ 
 1000 & 10 & 42.5 & 0.010 \\ 
 1000 & 100 & 41 & 0.0011 \\ 
 1000 & 1000 & 39 & 0.00012 \\ 
	\hline
 2000 & 1 & 52 & 0.16 \\ 
 2000 & 3 & 56 & 0.039 \\ 
 2000 & 10 & 57.5 & 0.011 \\ 
 2000 & 100 & 55.5 & 0.0011 \\ 
 2000 & 1000 & 52 & 0.00013 \\
\hline 
\end{tabular}
\end{center}
\caption{Numerical results of the fitting parameters $\Ne_\mathrm{PT}$ and $\calP_\curv^\mathrm{(peak)}$ for given $\Pi^2$ and $\D$.
}
\label{tab:1}
\end{table*}

\section{Parameters of interest for astrophysical PBH formation}
\label{sec:parameters}

In this section, we discuss which parameters should be chosen to generate the desired mass of PBHs consistently with other observed quantities. 
We then apply our result of the previous section to calculate the PBH mass function. We also show a GW spectrum predicted in our model.

\subsection{CMB constraints}

Since perturbations at the CMB scale exits the horizon before the waterfall phase transition, they are generated by the quantum fluctuation of the inflaton $\phi$.
The amplitude of curvature perturbations is therefore given by the standard textbook formula
\bae{ 
 	\calP_\curv = \frac{\HI^2}{8 \pi^2 \epsilon \Mpl^2},
}
where 
\bae{
 	\epsilon = \frac12 \lmk \Mpl^2 \frac{V_\phi}{V} \rmk^2
 	\simeq \frac{\Mpl^2}{2 \mu_1^2},
}
is a slow-roll parameter. 
This should be consistent with the observed one: 
\bae{
 	\calP_\curv (k_*) \simeq 2.1 \times 10^{-9},
	\label{COBE}
}
where $k_*$ ($= 0.05 \, {\rm Mpc}^{-1}$) represents the wavenumber at the pivot scale~\cite{Planck:2018vyg}. 
This implies
\bae{\label{eq: Lambda}
 	\Lambda \simeq 5.4 \times 10^{15} \GeV 
 	\lmk \frac{\Pi}{10} \rmk^{-1} 
	\lmk \frac{M}{\Mpl} \rmk 
 	\lmk \frac{\phi_\uc}{\Mpl} \rmk^{1/2},
}
or
\bae{
 	\HI \simeq 
 	7.0 \times 10^{9} \GeV 
	\lmk \frac{\mu_1}{10^5 \Mpl} \rmk^{-1}. 
}
In addition, using \eq{COBE} and \eq{chi1}, 
we obtain 
\bae{
 	\chi_2 
 	\simeq 
 	\frac12 \ln \lmk \frac{\Pi}{\D} \sqrt{2}{\pi} \frac{1}{2.1 \times 10^{-9}} \rmk 
 	\simeq
 	9.9 + \frac12 \ln \lmk \frac{\Pi}{\D} \rmk. 
}
Thus we expect $\chi_2 \sim 10$.

The spectral index in our model is provided by 
\bae{
	\ns\simeq
 	\eval{1+ 2\Mpl^2 \frac{V_{\phi \phi} }{V}}_{k_*} 
 	=
 	1 - \frac{4\Mpl^2}{\mu_2^2}. 
 	\label{ns}
}
We can consider $\mu_2 \simeq 10 \Mpl$ to explain the observed spectral index of $\ns = 0.9649 \pm 0.0042$~\cite{Planck:2018vyg}.

\subsection{Constraint from self-ordering scalar fields}

Our model has the global O$(\D)$ symmetry, which is spontaneously broken to O$(\D - 1)$ at the waterfall phase transition. 
For the case with $\D \le 4$, topologically non-trivial field configurations form after the \ac{SSB}. 
For higher $\D$, no stable topological defect exist, but the \ac{NG} modes are randomly distributed over the horizon scale and their gradient energy is proportional to the SSB scale squared. 
When those modes come into horizon at a late time, they tend to move in order to reduce the gradient energy~\cite{Turok:1991qq}. 
The dynamics of NG modes as well as topological defects can affect the CMB temperature anisotropy. 
Especially for the case with $\D =2$ and $3$, where cosmic strings (for $\D=2$) and monopoles (for $\D=3$) form at the phase transition, 
Ref.~\cite{Lopez-Eiguren:2017dmc} discussed a constraint on the SSB scale based on the Planck results such as 
\beae{
 	&M < 2.9 \times 10^{15} \GeV \qquad \text{for} \quad \D = 2, 
 	\\
 	&M < 6.4 \times 10^{15} \GeV \qquad \text{for} \quad \D = 3, 
}
We expect that the constraint for the case with large $\D$ is weaker by a factor of a few. 
Therefore we are interested in the case with $M \lesssim 10^{16} \GeV$. 
Note that this constraint can be evaded if one introduces a tiny but nonzero explicit O$(\D)$ breaking term in the Lagrangian because all NG modes are aligned by that term.

The dynamics of self-ordering scalar fields also emits gravitational waves~\cite{Krauss:1991qu,Jones-Smith:2007hib,Fenu:2009qf} (see also Ref.~\cite{Kamada:2015iga} in a different context). It has a flat spectrum in the frequency range of interest. Its amplitude is given by 
\bae{
 	\Omega_{\rm GW}^{\rm (SO)} = 1.2 \times 10^{-14} \times \frac{\Sigma_{\D}}{\D} \times \lmk \frac{M}{10^{16} \GeV} \rmk^4 \,,
 	\label{eq:GWSO}
}
where $\Sigma_{\D}$ is a numerical factor that approaches to unity for a large $\D$ limit. For example, $\Sigma_{\D} \simeq 4.1$ ($1.7$) for $\D = 4$ ($8$)~\cite{Figueroa:2020lvo}. 
This should be added to the gravitational wave signals from second-order effect from large curvature perturbations, which we will discuss shortly.

\subsection{PBH formation}

PBHs form when an overdense region with an ${\cal O}(1)$ density perturbation enters the horizon. 
A sizable amount of PBHs can form when~\cite{Carr:1975qj}%
\footnote{\label{footnote: PBH}
There are many studies in the literature to determine this condition in more detail~\cite{Shibata:1999zs,Harada:2015yda,Young:2019yug,Escriva:2019phb,Atal:2019erb} and take into account the effect of non-Gaussianity~\cite{Bullock:1996at,Ivanov:1997ia,Yokoyama:1998pt,Hidalgo:2007vk,Byrnes:2012yx,Bugaev:2013vba,Young:2015cyn,Nakama:2016gzw,Ando:2017veq,Franciolini:2018vbk,Atal:2018neu,Passaglia:2018ixg,Atal:2019cdz,Atal:2019erb,Yoo:2019pma,Taoso:2021uvl,Kitajima:2021fpq,Escriva:2022pnz}. 
See also Refs.~\cite{Yoo:2018kvb,Yoo:2019pma,Yoo:2020dkz,Kitajima:2021fpq} for a recent sophisticated evaluation of the PBH abundance in the peak theory. 
The detailed study for the PBH formation is beyond the scope of this paper. 
Because the PBH abundance has an exponential dependence on the amplitude of the density perturbations, 
our estimation is good enough to discuss if a sizable amount of PBH can form or not. 
}
\bae{
 	\calP_\curv^{\rm (peak)} \sim 0.01.
}
From the top-right panel of Fig.~\ref{fig: NPT and calPpeak}, 
the number of waterfall fields should be $\calO(1\text{--}10)$.

The perturbation of comoving wave length $\lambda$ enters the horizon when $a(t) \lambda (t) = 1/ H(t) \equiv 1/ H_{\rm HC}$. 
Assuming that this happens during the radiation dominated epoch, 
we obtain the corresponding e-folding number $N_{\rm PT}$ at which the relevant perturbation exits the horizon: 
\bae{
 	H_{\rm HC}^{-1} 
 	\sim
 	\frac{2}{\HI} \ee^{2 N_{\rm PT}} \lmk \frac{H_{\mathrm{inf}}}{H_{\rm RH}} \rmk^{1/3}, 
 	\label{H_HC}
}
where we use $t_{\rm end} \sim 1/ \HI$ and $t_{\rm RH} \sim 1/H_{\rm RH}$. 
The PBH mass is given by the total energy enclosed within the Hubble horizon 
at the horizon crossing: 
\bae{
 	M_{\rm BH} 
 	= \gamma \frac{4\pi}{H_{\rm HC}}, 
}
where $\gamma$ ($= {\cal O}(1)$) is a numerical constant~\cite{Carr:1975qj}. 
We thus obtain 
\bae{
	N_{\rm PT}
	&\simeq 19.5 
 	+ \frac{1}{2} \ln \lmk \frac{\HI}{10^{10} \GeV} \rmk 
 	+ \frac{1}{2} \ln \lmk \frac{M_{\rm BH}}{10^{20} \, {\rm g}} \rmk
 	+ \frac{1}{6} \ln \lmk \frac{H_{\rm RH}}{\HI} \rmk\, , \nonumber\\
  	&\simeq 19.4
 	- \frac{1}{2} \ln \lmk \frac{\mu_1}{10^5 \GeV} \rmk 
 	+ \frac{1}{2} \ln \lmk \frac{M_{\rm BH}}{10^{20} \, {\rm g}} \rmk
 	+ \frac{1}{6} \ln \lmk \frac{H_{\rm RH}}{\HI} \rmk\, , 
 	\label{N-PBH}
}
where we assume $\gamma = 0.2$. 
From the bottom panel of Fig.~\ref{fig: NPT and calPpeak},
we are interested in $\Pi \sim 10\text{--}20$.

We should take care of the tail of the spectrum for the curvature perturbations so that the spectrum around the CMB scale is not affected. 
Since the width of the spectrum $\Delta \Ne_1$ is of the same order with the peak frequency $\Ne_\mathrm{PT}$, the latter one cannot be arbitrary large. 
For the case of instantaneous reheating (i.e., $H_{\rm RH} = H_{\rm inf}$), 
we obtain $\Pi^2 \lesssim 250$ for $\phi_\uc / \sqrt{2} = M = 10^{16} \GeV$, which leads to $\Ne_\mathrm{PT}\lesssim 20$ and $M_{\rm PBH} \lesssim 2 \times 10^{25} \, {\rm g}$. 
For the case of the lowest possible reheating temperature of $H_{\rm RH} \sim 1 \MeV^2 / \Mpl$, 
we obtain $\Pi^2 \lesssim 200$ for $\phi_\uc / \sqrt{2} = M = 10^{16} \GeV$, which leads to $\Ne_\mathrm{PT}\lesssim 18$ and $M_{\rm PBH} \lesssim 1 \times 10^{27} \, {\rm g}$. 
The upper bounds on $\Pi$ or $\Ne_\mathrm{PT}$ do not change much for different values of $\phi_\uc$ and $M$ (with $\Pi$ fixed). Because of the relations of \eq{N-PBH} and $\Pi \equiv M \sqrt{\mu_1 \phi_\uc}/\Mpl^2$, 
a larger PBH mass can be generated 
for a smaller $M$ and $\phi_\uc$ under the same upper bound on $\Ne_\mathrm{PT}$.

\subsection{Summary of parameters of interest}

The parameters we are interested in are roughly given by 
\bae{
 	&\Pi \equiv 
 	10 \lmk \frac{\mu_1}{10^2 \Mpl} \rmk^{1/2} 
 	\lmk \frac{M}{\Mpl} \rmk 
 	\lmk \frac{\phi_\uc}{\Mpl} \rmk^{1/2} 
 	\sim 0.8 N_{\rm PT}, 
	\label{eq:cond1} \\
 	& \D \sim 0.6 \, \Pi \lmk \frac{{\cal P}_\curv^{\rm (peak)}}{0.01} \rmk^{-1},
 	\label{eq:cond2} \\
 	& \Lambda \simeq 5.4 \times 10^{15} \GeV 
 	\lmk \frac{\Pi}{10} \rmk^{-1} 
 	\lmk \frac{M}{\Mpl} \rmk 
 	\lmk \frac{\phi_\uc}{\Mpl} \rmk^{1/2}, 
 	\label{third} \\
 	& \mu_2 \simeq 10 \Mpl, \\
 	& M \lesssim  10^{16} \GeV, 
	\\
 	&\phi_\uc \lesssim \Mpl, \\
 	&\Ne_{\rm PT} \lesssim 18\text{--}20,
}
with $\Ne_{\rm PT}$ given by \eq{N-PBH}. 
The first two conditions, 
Eqs.~(\ref{eq:cond1}) and (\ref{eq:cond2}) are very rough estimation and are shown for illustrative purposes. 
More accurate conditions should be read by interpolating the numerical results in Table~\ref{tab:1}.
We added a sub-Planckian condition for the critical point: $\phi_\uc \lesssim \Mpl$. 
The parameters we have are 
$\mu_1$, $\mu_2$, $M$, $\phi_\uc$, $\Lambda$, and $\D$. 
Four of them, $\mu_1$, $\mu_2$, $\Lambda$, and $\D$, 
are (approximately) determined by the first four conditions. 
We have two free parameters, $M$ and $\phi_\uc$, 
which have the upper bounds. 

Using \eq{N-PBH} and the above conditions, Eqs.~(\ref{eq:cond1}) and (\ref{eq:cond2}), 
we roughly obtain 
\beq
 \D 
 &\sim& 9 \lmk \frac{{\cal P}_\curv^{\rm (peak)}}{0.01} \rmk^{-1} 
 \lmk 1 + \frac{1}{39} \lkk \ln \lmk \frac{M_{\rm BH}}{10^{20}\, {\rm g}} \rmk - 
 \ln \lmk \frac{\mu_1}{10^5 \GeV} \rmk
 + \frac{1}{3} \ln \lmk \frac{H_{\rm RH}}{H_{\rm inf}} \rmk
 \rkk \rmk \,. 
\eeq
This is the number of waterfall fields that is requied to generate PBHs with a mass of interest.

\subsection{Results}

We show an example that provides PBHs with a mass of order $10^{20}\, {\rm g}$, where we choose $\D = 5$, $\Pi^2 = 100$, and $\mu_2 = 10 \Mpl$ 
with $\phi_\uc / \sqrt{2} = M = 10^{16} \GeV$. 
Interporating the numerical results in Table~\ref{tab:1}, 
we obtain $\Ne_{\rm PT} \simeq 14$ and $\mathcal{P}_{\mathcal{R}}^{\rm (peak)} \simeq 0.014$ for this case.

\begin{figure}
    \centering
    \begin{tabular}{c}
        \begin{minipage}{0.5\hsize}
            \centering
            \includegraphics[width=0.95\hsize]{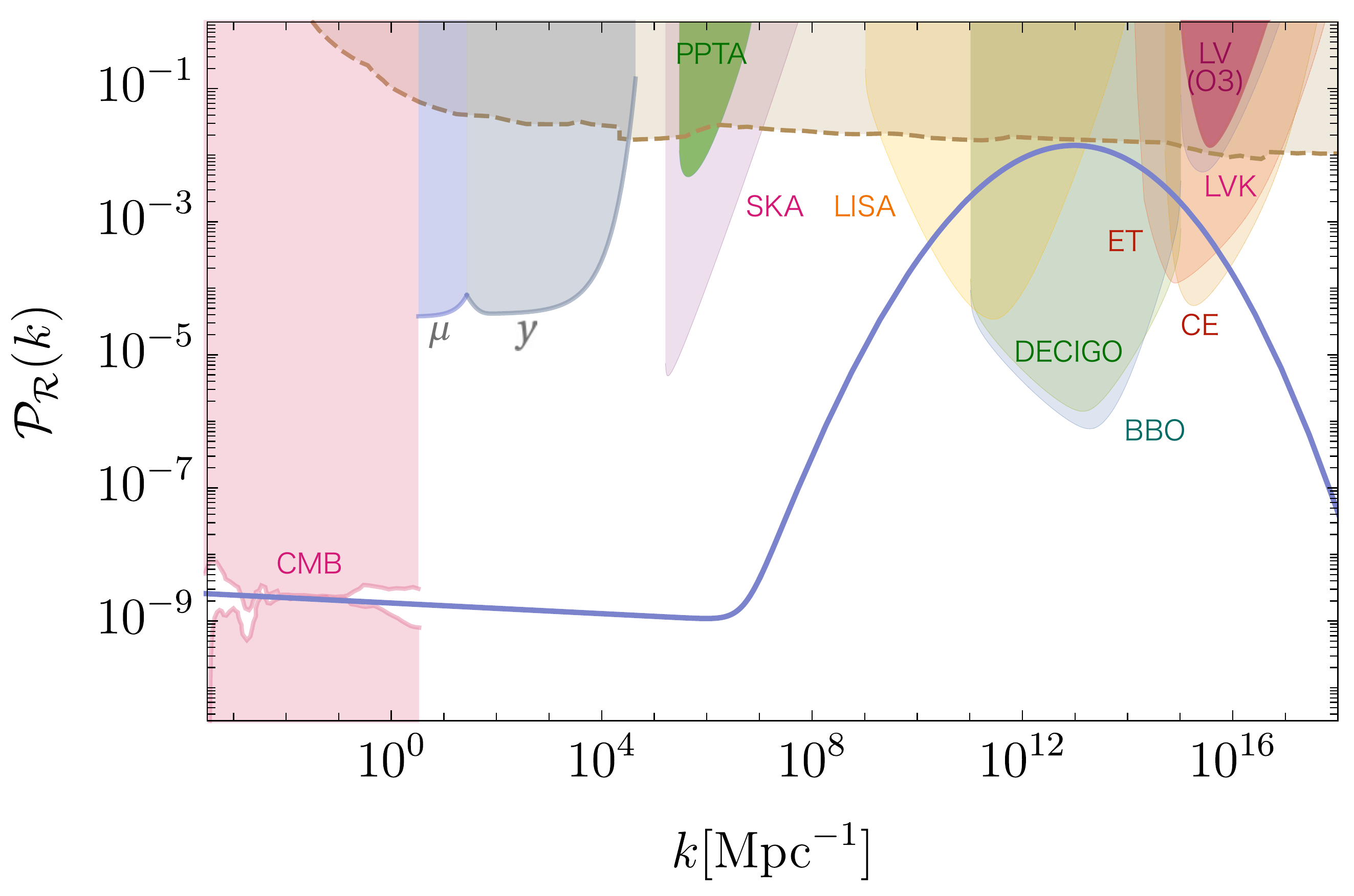}
        \end{minipage}
        \begin{minipage}{0.5\hsize}
            \centering
            \includegraphics[width=0.95\hsize]{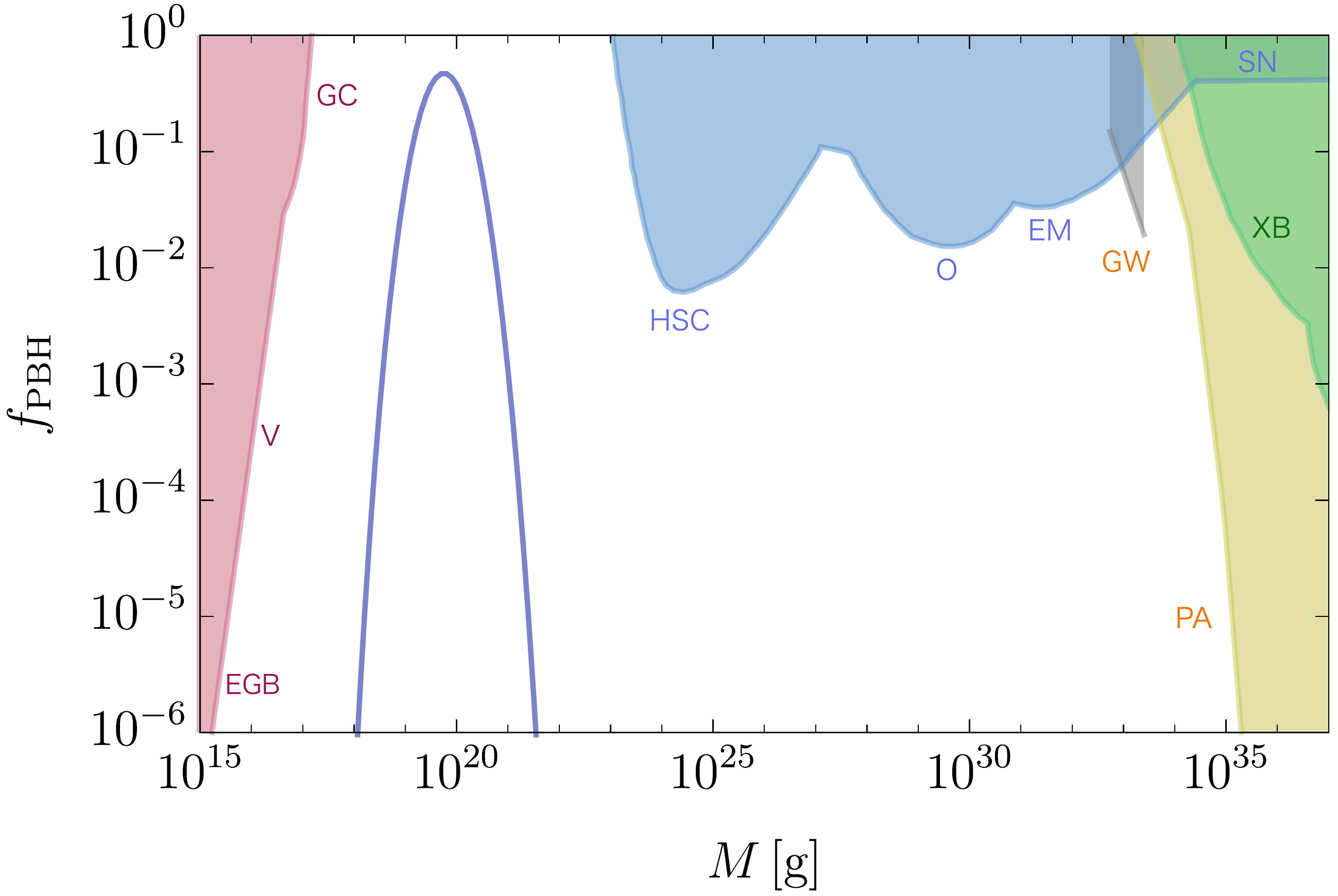}
        \end{minipage}
    \end{tabular}
    \caption{Spectra of curvature perturbations (left panel) and PBH mass function (right panel) for the case with $\D = 5$, $\Pi^2 = 100$, $\mu_2 = 10 \Mpl$, and $\phi_\uc / \sqrt{2} = M = 10^{16} \GeV$. The peak amplitude and frequency are determined by the interpolation for numerical results of stochastic-$\delta N$ formalism, whereas the shape of the spectrum is given by the analytic result \eq{eq:width0}. 
    }
    \label{fig:spectrum1}
\end{figure}

The spectrum of curvature perturbations is shown as the solid curve 
in the left panel of Fig.~\ref{fig:spectrum1}. The peak amplitude and frequency are determined by interpolating the numerical results in Table~\ref{tab:1}, 
whereas the shape of the spectrum is given by the analytic result \eq{eq:width0} to show a smooth curve. 
The shaded region above the dashed curve on the top of the figure is excluded by the overproduction of PBHs (see Ref.~\cite{Escriva:2022duf} and references therein). 
The red-shaded region in the left corner of the figure is excluded by the CMB observation~\cite{Nicholson:2009pi, Nicholson:2009zj, Bird:2010mp, Bringmann:2011ut}. 
The blue and green shaded region in the upper left corner is excluded by constraint for $\mu$ and $y$ distortions~\cite{Fixsen:1996nj, Chluba:2012we}.%
\footnote{See also Ref.~\cite{Bianchini:2022dqh} for a recent analysis.}
The constraint and future sensitivity curves for gravitational wave experiments are shown as the other dense and light-shaded regions, respectively (see Ref.~\cite{Schmitz:2020syl} and references therein). 

In the right panel of Fig.~\ref{fig:spectrum1}, 
we show the PBH mass function generated by the collapse of overdense regions. 
The peak mass for the PBH is about 
$5.3 \times 10^{19} \, {\rm g}$, which is within the window for the PBH DM. 
Note that, in order to keep the model parameters simple, we chose the threshold value $\delta_\uc$ ($\sim 0.3$) 
so that the total PBH energy density is equal to the observed DM density rather than fixing $\delta_\uc$ and fine-tuning the parameters. 
Hence this mass function should be understood as a schematic illustration. In fact, the threshold value with respect to the density contrast is known to be non-universal but depend on the profile of the overdensity and anyway one has to go beyond the Press--Schechter approach for a precise evaluation of the \ac{PBH} abundance. The non-Gaussianity of the curvature perturbation will also affect the \ac{PBH} abundance. We leave all these possible corrections for future works (see also the comments in footnote~\ref{footnote: PBH}).
The shaded regions are excluded by overproduction of PBHs (see, e.g., Ref.~\cite{Escriva:2022duf} and references therein for detail).

\begin{figure}
            \centering
            \includegraphics[width=0.75\hsize]{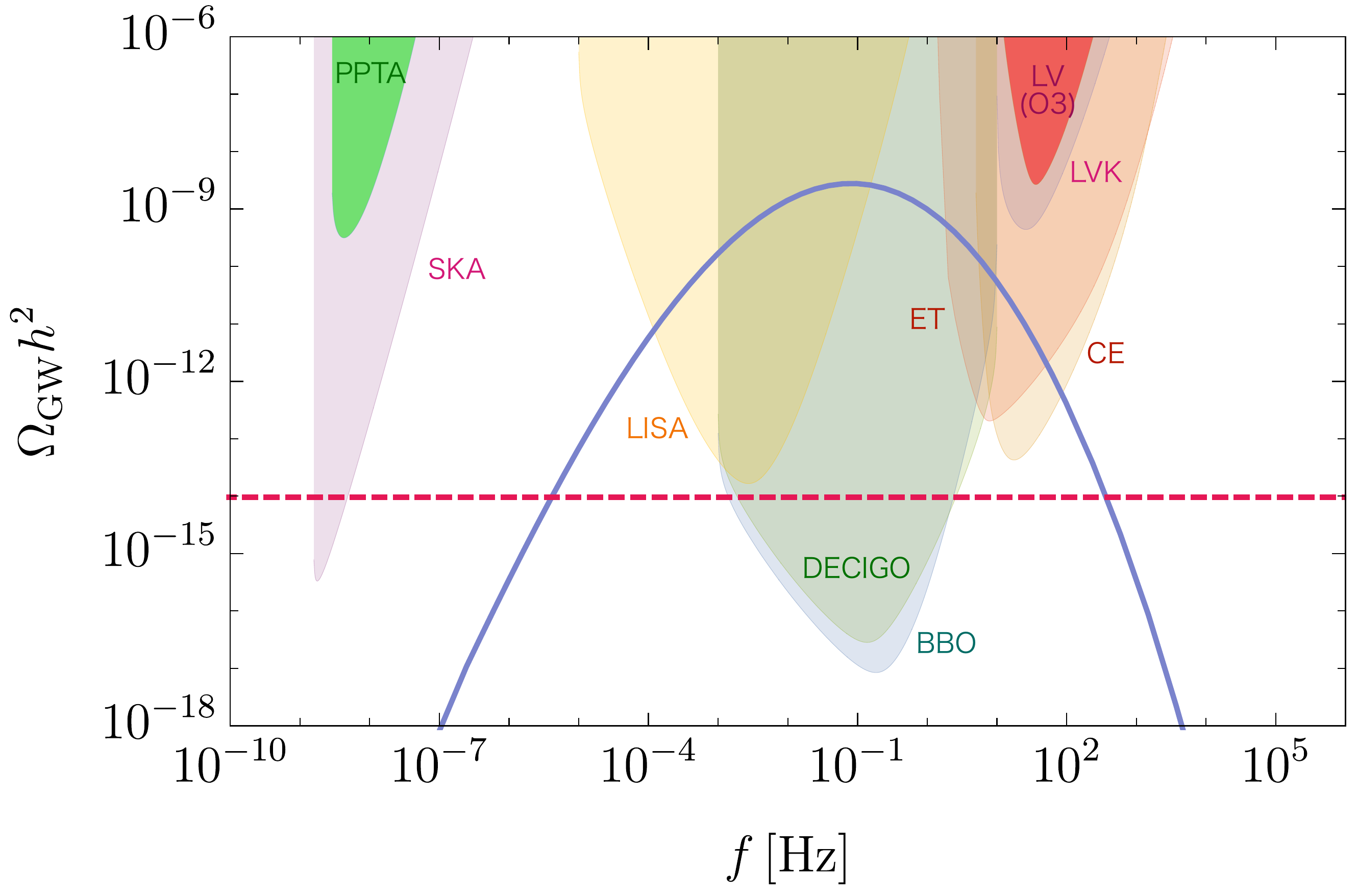}
    \caption{Spectra of stochastic gravitational waves generated by the dynamics of NG modes (red dashed line) and 
    the second-order effect from large curvature perturbations (solid blue curve). }
    \label{fig:spectrumGW}
\end{figure}

Fig.~\ref{fig:spectrumGW} shows the spectrum of stochastic gravitational waves predicted in our model. 
As we discussed around \eq{eq:GWSO}, gravitational waves are generated from the dynamics of re-alignment of NG modes when the modes enter into the horizon. In the figure, we plot the signal as the red dashed line for the case with $M = 10^{16} \GeV$ and $\D = 5$. We assume $\Sigma_{\D} = 4$. 
Stochastic gravitational waves are also generated by the second-order effect from large curvature perturbations. The solid blue curve represents the prediction for the case with the above parameter sets. 
The dense shaded regions are excluded by current experiments and the light-dense regions are future sensitivity curves for planned experiments.

\section{Discussion and conclusions}
\label{sec:conclusion}

We have analytically and numerically calculated the spectrum of curvature perturbations in the hybrid inflation model with O($\D$)-symmetric waterfall fields. 
The numerical simulation is based on the {\it stochastic}-$\delta \Ne$ algorithm proposed and developed in Refs.~\cite{Fujita:2013cna,Fujita:2014tja,Vennin:2015hra,Ando:2020fjm,Tada:2021zzj}. We formulate the stochastic dynamics under the O($\D$) symmetry and demonstrate that the resulting amplitude of curvature perturbations is reduced by a factor of order $\sqrt{\D}$ for a fixed peak frequency. 
This factor is important to generate PBHs within a desirable mass range from a collapse of the overdense region at the horizon entry. It has been observed that the PBHs with a mass of order $10^{20} \, {\rm g}$ 
can be generated consistently with other cosmological constraints if the number of waterfall fields is about $5$ for the case of instantaneous reheating. 
Such 
\acp{PBH} can explain all energy density of dark matter in the Universe. 
We have also calculated the spectrum of stochastic gravitational waves generated by the second-order effect from large curvature perturbations. The resulting GW signals can be observed in future GW experiments, including LISA.

The extended sector for the waterfall field is also motivated by avoiding the domain wall problem that arises in a model for a single waterfall field with $Z_2$ symmetry. 
The O$(\D)$ symmetry is spontaneously broken by the vacuum expectation value of waterfall fields after the waterfall phase transition. This implies that stochastic NG modes are produced, which results in a continuous re-ordering of NG modes after the phase transition for the case of $\D \ge 5$~\cite{Turok:1991qq}. The cosmological effect of those dynamics is qualitatively similar to the case with $\D = 2$, $3$, and $4$, where topologically non-trivial field configurations form after the spontaneous symmetry breaking. 
Those dynamics particularly affect the CMB spectrum and give an upper bound on the symmetry-breaking scale. 
Moreover, it generates GWs with a flat spectrum~\cite{Krauss:1991qu,Jones-Smith:2007hib,Fenu:2009qf}, which would be observed by future GW experiments as well as the pulsar-timing array experiments. 
This is an interesting smoking-gun signal of the present model.

One may consider that any global symmetries should not be exact to ensure consistency with quantum gravity. The \ac{NG} boson then has a small mass from an explicit symmetry-breaking term, and the stochastic modes are ordered by the mass term at a late time. 
In this case, the constraints from the re-ordering of NG modes may be evaded and the GW spectrum from the dynamics of self-ordering NG modes is suppressed at a small frequency.

Although we have specifically considered the case with global O($\D$) symmetry for waterfall fields, 
the qualitative results, such as the scaling of the amplitude of curvature perturbations as $\propto \D^{-1}$, are expected to hold for other symmetries such as SU($\D$). 
Moreover, one can consider the gauged O($\D$) or other symmetries instead of the global ones. 
The stochastic effect on the NG modes as well as the radial mode should be the same in a covariant gauge at least in the regime where the mass of the gauge boson is much smaller than the Hubble parameter during inflation (see, e.g., Refs.~\cite{Graham:2015rva,Sato:2022jya}). 
Since the vacuum expectation value of the waterfall fields is comparable to the Hubble parameter at the waterfall phase transition, our calculation can be justified at least for the case with a sufficiently small gauge coupling constant. 
The gauge field becomes heavier after the waterfall phase transition than the Hubble parameter. The stochastic structure of NG modes in the covariant gauge can be understood as the massive gauge fields in a unitary gauge. The heavy gauge fields are then supposed to decay into light fields. Therefore the constraint from CMB anisotropy and prediction of GW signals from NG modes are absent for $\D \ge 4$ in this case. Still, the GW signals from the second-order effect of large curvature perturbations are the prediction of the model.

\acknowledgments

YT is supported by JSPS KAKENHI Grant No. JP21K13918.
MY is supported by MEXT Leading Initiative for Excellent Young Researchers and JSPS KAKENHI Grants No. JP20H05851 and No. JP23K13092.

\bibliography{ref}
\end{document}